\documentclass[11pt]{article}
\usepackage[margin=2.5cm]{geometry}
\usepackage[T1]{fontenc}
\usepackage[utf8]{inputenc}
\usepackage{lmodern}
\usepackage{microtype}
\usepackage{graphicx}
\usepackage{booktabs}
\usepackage{tabularx}
\usepackage{array}
\usepackage{amsmath,amssymb}
\usepackage{textcomp}
\usepackage{caption}
\usepackage{placeins}
\usepackage{float}
\usepackage{pdflscape}
\usepackage[numbers,sort&compress]{natbib}
\usepackage[colorlinks=true,linkcolor=black,citecolor=black,urlcolor=black,filecolor=black,hypertexnames=false]{hyperref}
\usepackage{lineno}
\usepackage{xcolor}

\graphicspath{{media/}}
\newcolumntype{L}[1]{>{\raggedright\arraybackslash}p{#1}}
\newcolumntype{Y}{>{\raggedright\arraybackslash}X}
\renewcommand{\topfraction}{0.90}
\renewcommand{\bottomfraction}{0.80}
\renewcommand{\textfraction}{0.10}
\renewcommand{\floatpagefraction}{0.75}
\nolinenumbers
\begin{document}
\begin{center}
{\LARGE\bfseries Inclined surfaces bias bacterial swarming expansion and edge organization\par}
\vspace{1em}
Pengxi Gu$^{1}$, Zixiang Lin$^{1}$, Hui Ren$^{1}$, Jin Zhu$^{2}$, Anqi Li$^{1}$, Omar Said$^{3}$, Weijie Chen$^{4}$ and Zijie Qu$^{1,*}$\\[0.75em]
$^{1}$Global College, Shanghai Jiao Tong University, Shanghai, 200240, China.\\
$^{2}$School of Physics, Georgia Institute of Technology, Atlanta, GA 30332, USA.\\
$^{3}$Department of Mechanical Engineering, Johns Hopkins University, Baltimore, MD 21218, USA.\\
$^{4}$Intelligent Medicine Institute, Shanghai Medical College, Fudan University, Shanghai, 200433, China.\\[0.75em]
$^{*}$Correspondence: Zijie Qu (zijie.qu@sjtu.edu.cn)
\end{center}
\vspace{1em}
\noindent\textbf{Abstract}\quad

Bacterial swarming is the collective expansion of dense populations of
motile cells across hydrated soft surfaces, yet gravity is usually
treated as a fixed background condition in standard horizontal assays.
We used surface inclination to examine how agar-surface orientation
relative to gravity affects the expansion of \textit{Enterobacter}
sp. SM3 swarms. Colony expansion depended non-monotonically on
inclination, with the strongest spreading at intermediate inclinations
and weaker expansion under shallow or near-inverted conditions. Paired
up-facing acute and down-facing obtuse configurations with the same
tilt magnitude showed different expansion responses, suggesting that
swarm expansion depended on both tilt magnitude and which side of the
agar surface faced upward relative to gravity. Inclination also produced downhill-biased colony morphologies.
PIV analysis showed generally lower mean activity under inclination than in the flat
references, with a non-monotonic angular response and inward increases in
speed and spatial velocity-correlation length. The
downhill tip showed slower motion, shorter spatial correlations and smaller
vortex cores than the upper region. To interpret the macroscopic trends, we extended a two-phase
thin-film model by introducing an effective gravity coupling projected
normal and tangential to the agar surface. Simulations suggest that the
surface-tangential component mainly drives lateral redistribution,
whereas the surface-normal component modifies film pressure and thereby
pressure-dependent water replenishment, helping distinguish up-facing
from down-facing configurations.

\vspace{1em}


\section*{Introduction}

Bacterial swarming is a collective mode of surface migration in which dense populations of motile cells spread over semi-solid substrates \cite{harshey_surface_2003,kearns_field_2010,partridge_surveying_2022}. Unlike swimming in bulk liquid, swarming occurs within a thin surface-bound fluid layer and
generates coordinated cell motion, transient clusters, vortices and collective flows across microscopic and mesoscopic scales \cite{darnton_dynamics_2010,zhang_collective_2010,beer_statphys_2019,aranson_bacterial_2022}. Swarming dynamics are shaped by cell growth, flagellar motility, cell--cell alignment, wetting, surfactant production and substrate friction \cite{kearns_field_2010,copeland_bacterial_2009,darnton_dynamics_2010,aranson_bacterial_2022}. At the colony scale, swarming forms expanding circular, dendritic or tendril-like patterns, which have been modeled through combinations of thin-film expansion theories, cell growth, surfactant production and wetting mechanisms \cite{kearns_field_2010,trinschek_front_2018,srinivasan_multiphase_2019,PhysRevE.101.012407,kumar_generalized_2025}. Bacterial swarming therefore provides a natural active matter system for studying collective self-propelled motion across multiple scales at interfaces \cite{marchetti_hydrodynamics_2013,aranson_bacterial_2022}.

Recent work has made clear that the physical environment is an integral part of swarming. Swarm expansion depends on hydration and fluid recruitment from agar \cite{wu_microbubbles_2011,wu_water_2012,ping_osmotic_2014}, surface tension and surfactant-mediated wetting \cite{angelini_bacillus_2009,fauvart_surface_2012,ke_water_2015,trinschek_front_2018}, and the coupling between cells and aqueous films \cite{wu_microbubbles_2011,srinivasan_multiphase_2019,kumar_generalized_2025}. Among these environmental factors, gravity has received little attention. Standard experiments are usually performed on horizontal agar plates, where gravity is treated as a fixed and spatially neutral background condition. Yet microbial organizations are influenced by gravity when cells grow in weakly constrained or geometrically biased environments. A classic soft-agar study showed that the direction of gravity altered the colonial morphology of staphylococci \cite{wilkins_effect_1969}. More recent biofilm and microfluidic studies have shown that gravity and sedimentation affect bacterial accumulation, motility and early biofilm formation at solid--liquid interfaces \cite{marra_microfluidics_2025,malea_contribution_2025}. In active matter more broadly, gravity is increasingly recognized as a physical field that affects sedimentation, convection, particle alignment and collective organization \cite{volpe_active_2022}. Although several observations suggest that gravity affects microbial communities on inclined or biased surfaces \cite{ke_water_2015,hennes_active_2017,malea_contribution_2025}, the role of gravity in swarming remains unclear. 

To address this question, we used \textit{Enterobacter} sp. SM3 as a model swarming bacterium.  SM3 was originally isolated from the mouse gut, a setting in which microbial migration usually occurs on curved and non-horizontal hydrated surfaces. SM3 forms dense swarming colonies with strong microscopic collective motion \cite{chen_confinement_2021}. On horizontal agar, SM3 typically develops a nearly round colony outline, which provides a simple reference state for detecting inclination-induced changes in expansion and morphology.  

We examined SM3 swarming across different inclinations using
colony-scale expansion measurements, directional morphology analysis
and PIV (Particle Image Velocimetry) analysis at five locations around the colony periphery. We compared
velocity profiles with inward distance, regional activity and the
spatial organization of collective motion. Inclination produced a non-monotonic expansion response and
downhill-biased colony shapes. To interpret the macroscopic trends, we extended a two-phase thin-film
model with pressure-dependent Darcy-type replenishment by introducing
an effective gravity coupling projected normal and tangential to the agar
surface. In this model, the surface-tangential component biases lateral
film transport, whereas the surface-normal component modifies the film
pressure and thereby pressure-dependent agar-to-swarm water exchange. These results suggest that gravity acts as a directional physical bias that couples substrate water exchange, surface-film transport, edge motion and colony-scale pattern formation.

\section*{Results}
\subsection*{Inclination changes swarm expansion non-monotonically over time}
\begin{figure}[htbp]
\centering
\includegraphics[width=0.99\textwidth]{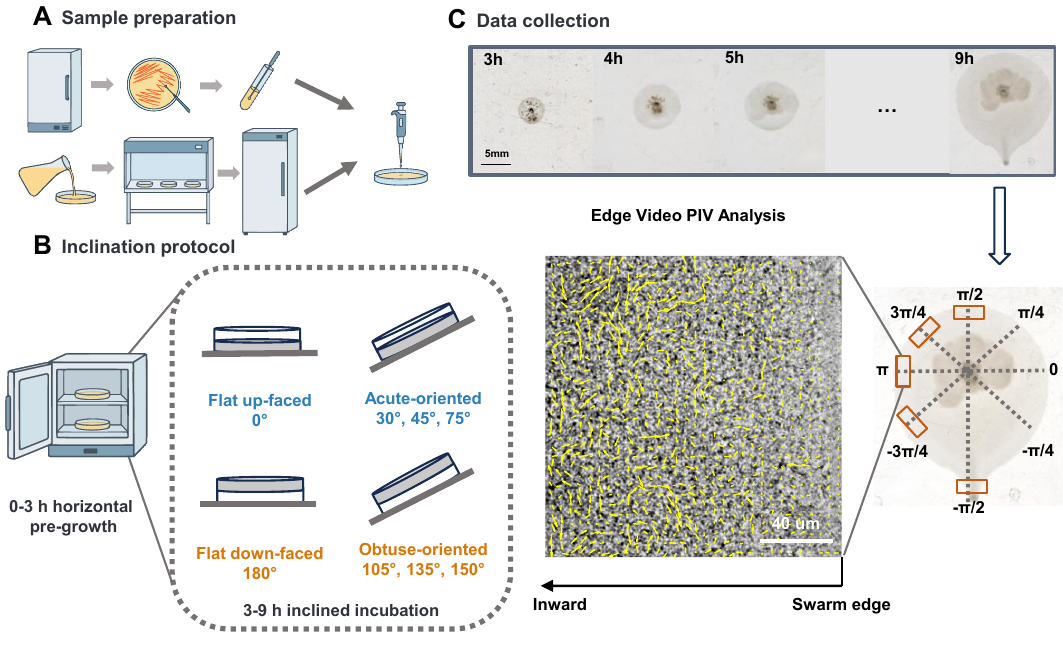}
\caption{
Experimental workflow, inclination protocol, and edge-resolved imaging
strategy for SM3 swarming.
\textbf{A}, Sample preparation. Cells were recovered from frozen stock,
streaked on LB agar and grown in liquid LB medium. Semi-solid 0.5\% LB
agar plates were dried for 10~min in a laminar-flow hood and stored at
\(4~^\circ\mathrm{C}\) for 7 days to standardize agar hydration. A
\(1.85~\mu\mathrm{L}\) droplet was inoculated at the center of each
prepared plate.
\textbf{B}, Inclination protocol. After 3~h of horizontal pre-growth,
plates were transferred to one of eight agar-surface orientations:
flat up-facing (\(0^\circ\)), up-facing tilted
(\(30^\circ\), \(45^\circ\), \(75^\circ\)), down-facing tilted
(\(105^\circ\), \(135^\circ\), \(150^\circ\)), and flat down-facing
(\(180^\circ\)) conditions.
\textbf{C}, Macroscopic and microscopic data collection. Colony images
were acquired from 3 to 9~h. Microscopic videos were recorded at five
gravity-defined sectors: upper (\(\pi/2\)), upper diagonal
(\(3\pi/4\)), side (\(\pi\)), lower diagonal (\(-3\pi/4\)) and
leading tip (\(-\pi/2\)). The representative microscopic field shows
PIV vectors overlaid on the acquired image, with the swarm edge at the
right and the inward direction pointing left. Complete
\(1500\times1500\)-px fields were used for the principal microscopic
comparisons (Figs.~\ref{fig:micro_activity_sector},\ref{fig:micro_mesoscopic_state}).
Scale bars, 5~mm in the colony images and 40~\(\mu\mathrm m\) in the
microscopic field. The sampled inward coordinate is defined in Materials and Methods.
}
\label{fig:whole_setup}
\end{figure}

Small droplets of SM3 culture were inoculated at the center of semi-solid 0.5\% LB agar plates. After 3~h of horizontal pre-growth at \(30~^\circ\mathrm{C}\) and \(90\%\) RH, plates were transferred to eight prescribed agar-surface configurations relative to gravity, including \(0^\circ\), \(30^\circ\), \(45^\circ\), \(75^\circ\), \(105^\circ\), \(135^\circ\), \(150^\circ\) and \(180^\circ\) (Fig.~\ref{fig:whole_setup}). Colonies were imaged once per hour from 3 to 9~h, and colony area and directional radius were extracted from the time-lapse images. Detailed procedures are provided in Materials and Methods.

\begin{figure}[htbp]
\centering
\includegraphics[width=0.95\textwidth]{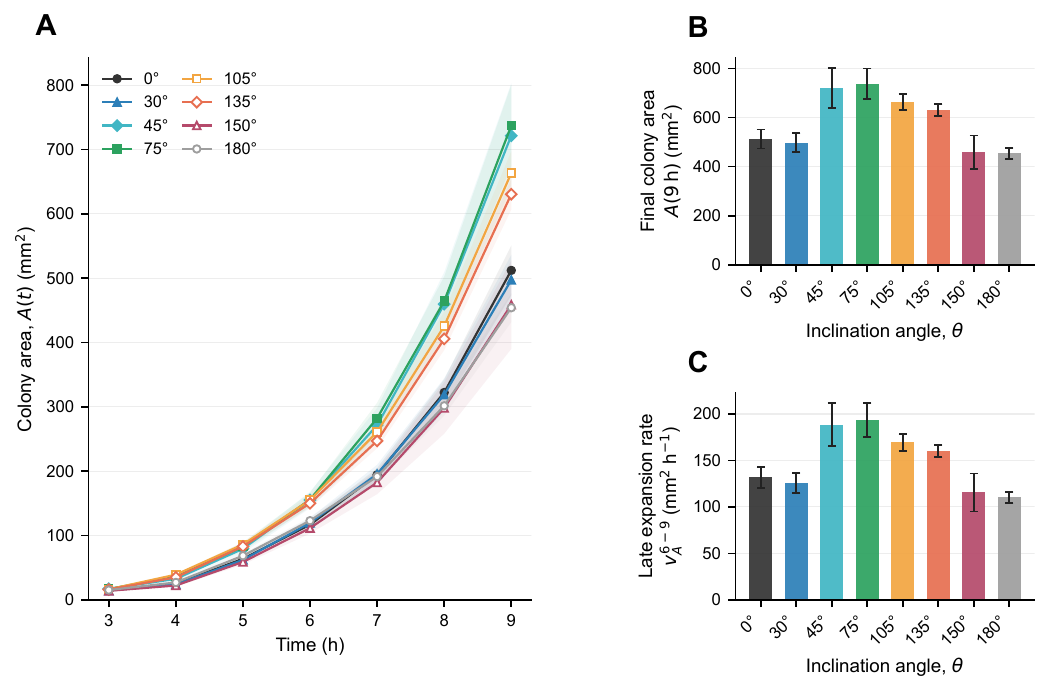}
\caption{
Surface inclination produces a time-accumulated, non-monotonic change in SM3 swarm expansion, with the strongest late-stage spreading at intermediate inclinations.
\textbf{A}, Mean colony area \(A(t)\) from 3 to 9~h for the eight agar-surface orientations. Inclination-dependent differences accumulated progressively during expansion.
\textbf{B}, Final colony area \(A(9~\mathrm{h})\).
\textbf{C}, Late-stage area expansion rate,
\(v_A^{6-9}=[A(9~\mathrm{h})-A(6~\mathrm{h})]/3~\mathrm{h}\).
Intermediate inclinations, particularly \(45^\circ\) and \(75^\circ\), showed greater final expansion and late-stage expansion rates than shallow or near-inverted conditions.
Data are shown as mean \(\pm\) s.e.m.; \(n=15\) independent colonies per inclination.
}
\label{fig:macro_expansion}
\end{figure}

Inclination-dependent differences in swarm expansion emerged progressively over time (Fig.~\ref{fig:macro_expansion}A). Colony areas were relatively similar across conditions at the beginning, but the trajectories separated as expansion proceeded. This divergence arose primarily during the later stage, when colonies at intermediate inclinations (\(45^\circ\), \(75^\circ\), \(105^\circ\), \(135^\circ\)) continued to expand rapidly, whereas those under low-inclination (\(0^\circ\), \(30^\circ\)) and near-inverted (\(150^\circ\), \(180^\circ\)) conditions accumulated area more slowly.

Swarm expansion varied non-monotonically with inclination and was strongest at intermediate angles. As shown in Fig.~\ref{fig:macro_expansion}B, relative to the horizontal \(0^\circ\) condition, final area was slightly lower at \(30^\circ\), increased markedly at intermediate acute inclinations, remained relatively large at \(105^\circ\) and \(135^\circ\), and declined again at \(150^\circ\) and \(180^\circ\). The largest final colonies occurred at \(75^\circ\) and \(45^\circ\), whereas \(150^\circ\) and \(180^\circ\) produced the smallest. Figure~\ref{fig:macro_expansion}C shows that the late-stage expansion rate followed the same overall angular dependence, with faster expansion at intermediate inclinations and slower expansion at \(30^\circ\), \(150^\circ\) and \(180^\circ\). Both measurements indicate stronger expansion at intermediate inclinations and weaker expansion at lower or near-inverted inclinations.

Comparison of supplementary angle pairs showed expansion dependence on the surface configuration. In our geometry (Fig.~\ref{fig:whole_setup}B), acute angles corresponded to up-facing tilted plates, whereas their supplementary obtuse angles corresponded to down-facing tilted plates with the same tilt magnitude. If inclination acted only as a scalar perturbation, each supplementary pair would be expected to produce similar expansion. In fact, paired up-facing and down-facing configurations gave different mean responses. The up-facing acute configurations tended to produce larger final colonies than their paired down-facing obtuse configurations, and the late-stage expansion rates showed a similar tendency. This tendency was seen for all pairs: \(30^\circ/150^\circ\), \(45^\circ/135^\circ\), \(75^\circ/105^\circ\) and \(0^\circ/180^\circ\) (Fig.~\ref{fig:macro_expansion}B,C). We therefore treat the supplementary-angle comparison as evidence that swarm expansion depends not only on tilt magnitude, but also on the up-facing or down-facing configuration of the agar surface.

\subsection*{Inclination biases swarm morphology downhill}

\begin{figure}[tp]
\centering
\includegraphics[width=0.87\textwidth]{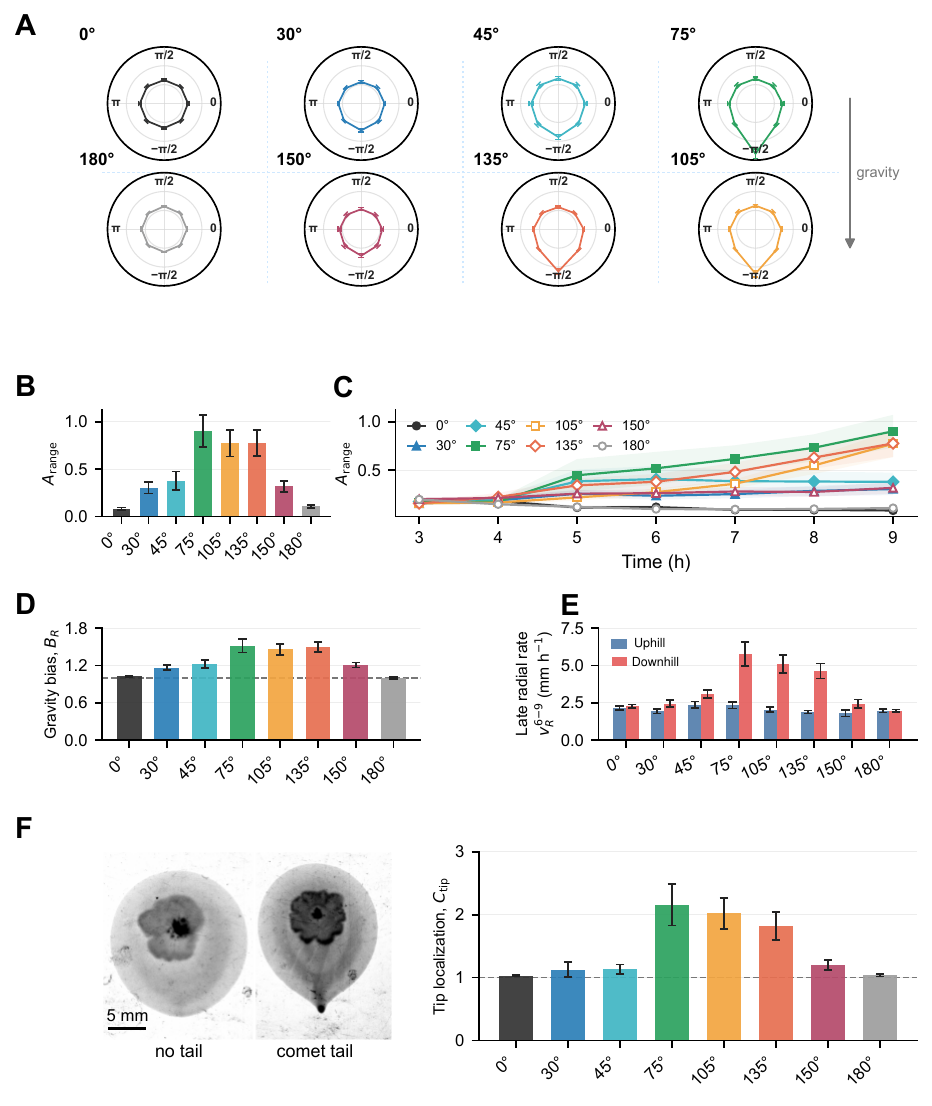}
\caption{
Inclination progressively generates downhill-biased swarm anisotropy
and can localize late-stage expansion at the downhill tip.
\textbf{A}, Final colony polar-radius profiles at 9~h, shown using a
common angular convention. The downhill direction is \(-\pi/2\), and
supplementary angle pairs are aligned vertically for comparison.
\textbf{B}, Final range-over-mean anisotropy
\(A_{\mathrm{range}}=(R_{\max}-R_{\min})/\langle R\rangle\) at 9~h.
\textbf{C}, Time evolution of \(A_{\mathrm{range}}\) from 3 to 9~h.
\textbf{D}, Downhill bias
\(B_R=\langle R_{-3\pi/4},R_{-\pi/2},R_{-\pi/4}\rangle/
\langle R_{\pi/4},R_{\pi/2},R_{3\pi/4}\rangle\); the dashed line at
\(B_R=1\) denotes equal upper- and lower-side extension.
\textbf{E}, Late-stage directional radial expansion rate
\(v_R^{6-9}\) along the uphill (\(\pi/2\)) and downhill
(\(-\pi/2\)) directions.
\textbf{F}, Representative colonies without and with a comet-like
downhill tail, together with the tip-localization index
\(C_{\mathrm{tip}}=v_R^{6-9}(-\pi/2)/
\langle v_R^{6-9}(-3\pi/4),v_R^{6-9}(-\pi/4)\rangle\). The dashed
line at \(C_{\mathrm{tip}}=1\) denotes equal tip and flank expansion.
Data are shown as mean \(\pm\) s.e.m.; \(n=15\) independent colonies
per inclination.
}
\label{fig:morphology}
\end{figure}

The inclination-dependent changes in expansion were accompanied by a directional reshaping of the swarm. Final polar-radius profiles showed that the \(0^\circ\) and \(180^\circ\) reference colonies remained nearly circular, whereas inclined colonies developed directionally biased outlines, most prominently at intermediate inclinations (Fig.~\ref{fig:morphology}A).

Overall shape anisotropy increased during inclined spreading (Fig.~\ref{fig:morphology}B,C). At 9~h, the range-over-mean anisotropy remained close to 0.1 in the two flat reference conditions, increased under inclination, and reached \(0.901\pm0.170\) at \(75^\circ\). It remained elevated at \(105^\circ\) and \(135^\circ\) before decreasing again at \(150^\circ\) (Fig.~\ref{fig:morphology}B). All conditions were only weakly anisotropic at 3~h, whereas \(75^\circ\), \(105^\circ\) and \(135^\circ\) progressively diverged from the flat references during subsequent spreading (Fig.~\ref{fig:morphology}C). 

This shape asymmetry was directed toward the downhill side of the colony. The downhill bias \(B_R\) remained near unity in the \(0^\circ\) and \(180^\circ\) references but increased to approximately \(1.46\)--\(1.52\) at \(75^\circ\)--\(135^\circ\) (Fig.~\ref{fig:morphology}D). Directional expansion rates over 6--9~h showed the same polarity, with the downhill front advancing more rapidly than the uphill front across these intermediate inclinations (Fig.~\ref{fig:morphology}E). The inclination-induced anisotropy therefore reflected preferential downhill extension rather than uniform radial enlargement.

In some inclined colonies, the downhill bias became further localized into a comet-like tail. 
To quantify this localization, we defined the tip-localization index \(C_{\mathrm{tip}}\), as described in the Methods,
which compares late-stage expansion at the downhill tip with that at the two adjacent flanks.
\(C_{\mathrm{tip}}\) remained close to unity in the \(0^\circ\) and \(180^\circ\) references but increased at intermediate inclinations, reaching \(2.16\pm0.33\) at \(75^\circ\) (Fig.~\ref{fig:morphology}F). Localized front release offers a possible explanation for these extensions \cite{hennes_active_2017}, as considered in the Discussion.

\subsection*{Inclination reorganizes collective swarm motion}

\begin{figure}[tp]
\centering
\includegraphics[width=.98\textwidth,height=.80\textheight,keepaspectratio]{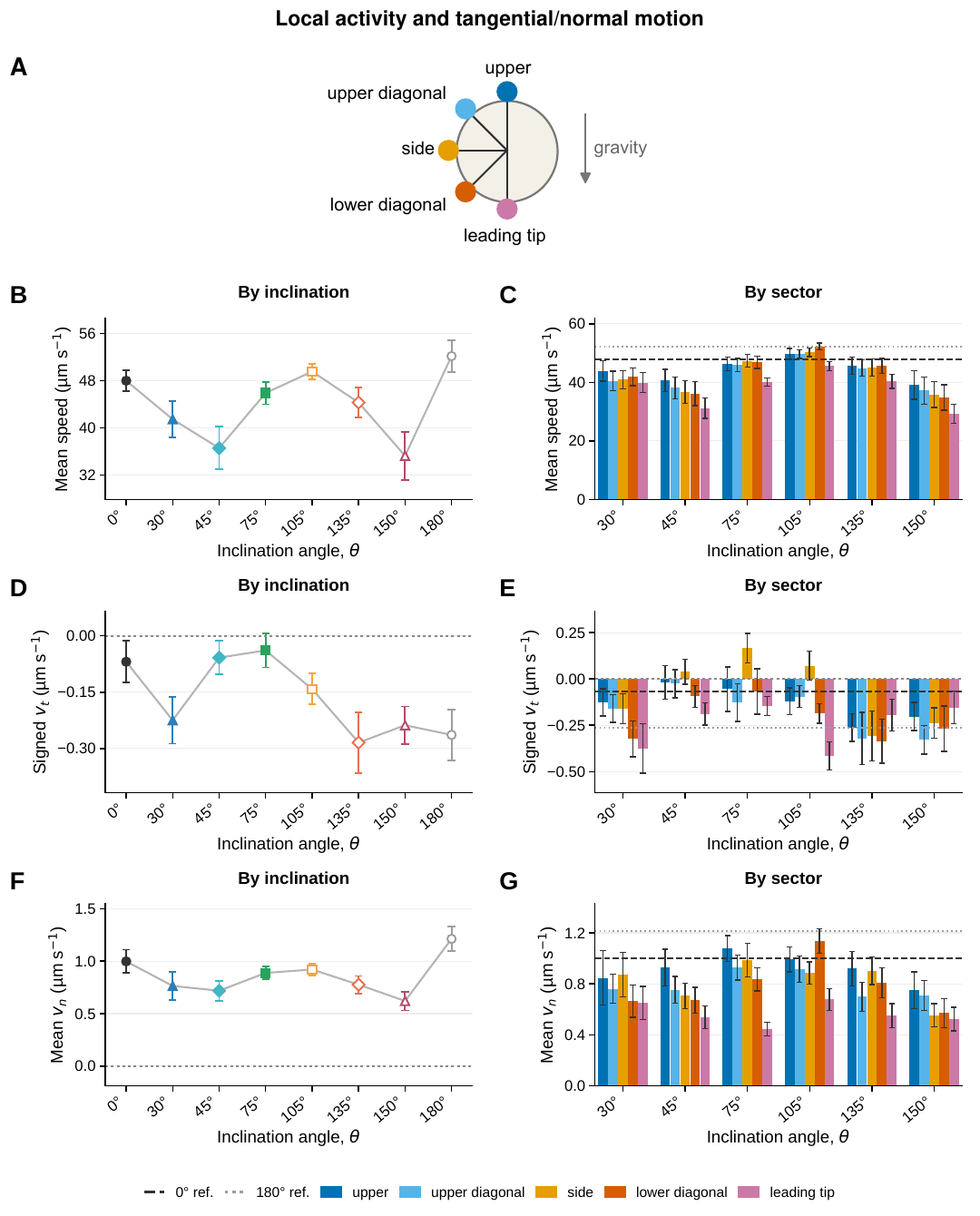}
\caption{Local activity and tangential/normal motion around the swarm periphery.
\textbf{A}, Five sampling sectors and their color code; the arrow indicates
projected gravity. \textbf{B,C}, Mean speed. \textbf{D,E}, Signed tangential
velocity. \textbf{F,G}, Mean normal velocity. B--G summarize 10.8--140.8~\(\mu\mathrm m\) inward from the
swarm edge. Quantitative panels show inclination means on the left and
sector means within tilted conditions on the right. Data are plate-level
mean \(\pm\) s.e.m.; 15 plates per inclination. Repeated recordings
were averaged within plate-sector; unavailable sectors were not imputed.
Dashed and dotted lines in C,E,G denote flat-reference means. The sign of
\(v_t\) denotes the image-coordinate tangent. The inward coordinate is
specified in Materials and Methods; speed profiles are shown in
Supplementary Fig.~\ref{fig:si_activity_support}A,B.}
\label{fig:micro_activity_sector}
\end{figure}

We used PIV (Particle Image Velocimetry) \cite{thielicke_pivlab_2021}
to examine collective motion as a function of inward distance from the
swarm edge, inclination angle and peripheral sector
(Figs.~\ref{fig:whole_setup}C,\ref{fig:micro_activity_sector}A).
Mean speed increased rapidly over the first few tens of micrometres
and approached a plateau farther inward
(Supplementary Fig.~\ref{fig:si_activity_support}A,B).
These two portions of the profile motivated equal-width outer and inner
intervals of 100--400~px (10.8--43.3~\(\mu\mathrm m\)) and
1000--1300~px (108.3--140.8~\(\mu\mathrm m\)), respectively.
Comparisons across inclination and sector used distance-averaged
measurements over the common retained interval of 100--1300~px
(10.8--140.8~\(\mu\mathrm m\)).

Mean activity was generally lower on tilted plates than in the flat
references. Mean speed varied non-monotonically with inclination,
decreasing at low inclinations, recovering at intermediate inclinations
and declining again toward near-inversion
(Fig.~\ref{fig:micro_activity_sector}B).
At every tilted inclination, speed decreased from the upper sector to
the leading tip (Fig.~\ref{fig:micro_activity_sector}C).
This decrease was relatively gentle near the middle of the inclination
range and steeper at lower and near-inverted inclinations
(Supplementary Fig.~\ref{fig:si_regional_support}A).
Root-mean-square (RMS) vorticity showed a similar activity pattern
(Supplementary Fig.~\ref{fig:si_activity_support}C,D).

We next examined the mean velocity components tangential and normal
to the local swarm edge, \(v_t\) and \(v_n\), to characterize net
directional motion within the sampled flow
(Supplementary Eq.~\ref{eq:si_native_mean}).
The signed tangential mean was more negative in the outer interval
(\(-0.297\pm0.037~\mu\mathrm m\,\mathrm s^{-1}\)) than in the inner
interval (\(-0.102\pm0.032~\mu\mathrm m\,\mathrm s^{-1}\)).
The negative tangential bias weakened toward intermediate inclinations
and became more pronounced again at higher inclinations
(Fig.~\ref{fig:micro_activity_sector}D).
Around the periphery, \(v_t\) was more negative at the leading tip than
at the side (Fig.~\ref{fig:micro_activity_sector}E).
Mean normal velocity remained positive and was lower at the leading tip
than in the upper sector at every tilted inclination
(Fig.~\ref{fig:micro_activity_sector}F,G).

\begin{figure}[htbp]
\centering
\includegraphics[width=.98\textwidth]{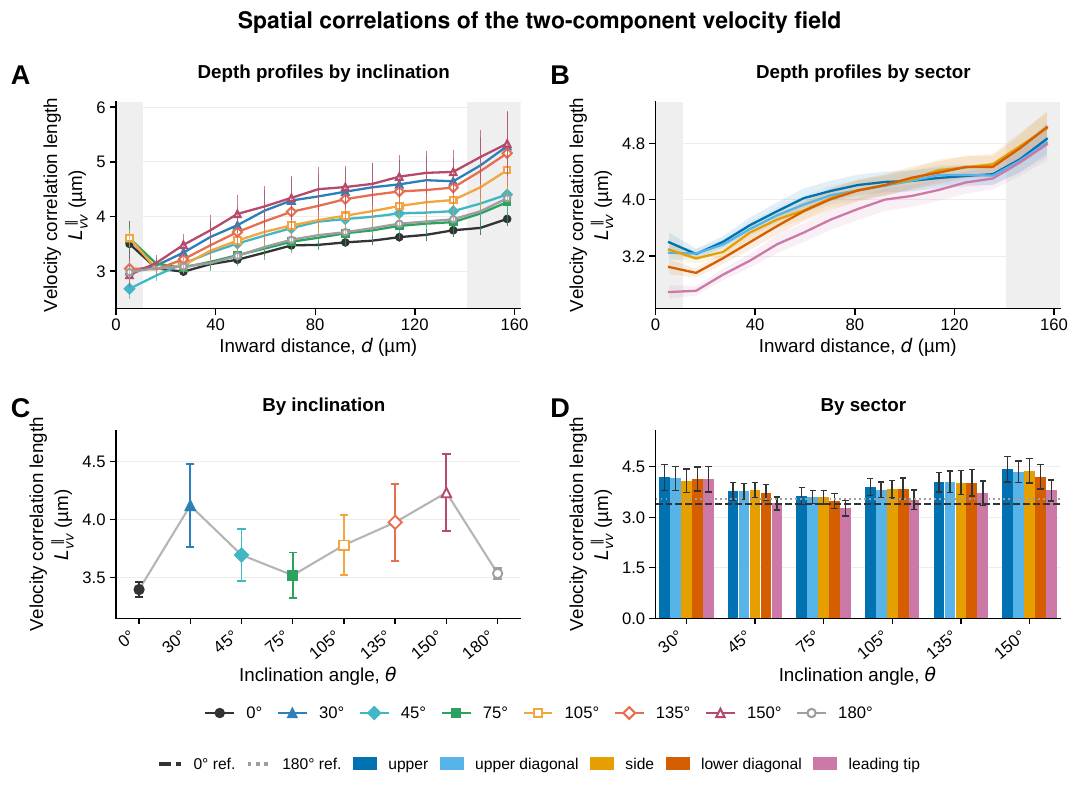}
\caption{Spatial velocity correlations increase inward and differ around the colony periphery.
\textbf{A,B}, Depth profiles of the full-vector correlation length,
\(L_{vv}^{\parallel}\), grouped by inclination and sector.
\textbf{C,D}, The same quantity averaged over 10.8--140.8~\(\mu\mathrm m\),
shown by inclination and by sector within each tilted condition.
The correlation uses the dot product of the two-component velocity
at points separated along the local edge tangent; \(d\) measures position
inward from the swarm edge. Gray boundary strips are excluded from
interval comparisons. Data are plate-level mean \(\pm\) s.e.m., with
15 independent plates per inclination and observed sectors retained
without imputation. Dashed and dotted lines in D denote flat-reference
means. The direction of the correlation separation is along the edge.}
\label{fig:micro_mesoscopic_state}
\end{figure}

At the mesoscopic scale, the spatial velocity-correlation length,
\(L_{vv}^{\parallel}\), increased inward across inclinations
(Fig.~\ref{fig:micro_mesoscopic_state}A).
Correlations were shorter near the leading tip, and sector differences
became smaller farther inward (Fig.~\ref{fig:micro_mesoscopic_state}B).
Averaged over the common 10.8--140.8~\(\mu\mathrm m\) interval, the
correlation length decreased from \(3.99\pm0.13~\mu\mathrm m\) in the
upper sector to \(3.66\pm0.13~\mu\mathrm m\) at the tip.
The sector effect was significant, while no overall inclination effect
was detected (Fig.~\ref{fig:micro_mesoscopic_state}C,D).

Vortices, identified through their rotating cores using the
\(\Gamma_1\)--\(\Gamma_2\) method \cite{graftieaux_combining_2001},
provided a complementary description of mesoscopic organization.
Their equivalent core diameter increased inward and was smaller at
the tip than in the upper sector (\(7.15\) versus
\(7.47~\mu\mathrm m\)), while vortex density was higher at the tip
(\(1.93\) versus \(1.71\) per \(1000~\mu\mathrm m^2\)).
Neither descriptor showed an overall inclination effect
(Supplementary Fig.~\ref{fig:si_vortex_support}).

Overall, inclination affected mean activity, while spatial organization
varied with inward distance and peripheral location. The downhill tip
had slower motion, shorter correlations and smaller vortex cores at
higher density.

\FloatBarrier

\subsection*{Gravity-projected simulations qualitatively recapitulate macroscopic inclination responses}

To test whether the macroscopic features could emerge from
gravity-biased transport, we extended the two-phase swarming expansion
model of Kumar and Pushpavanam~\cite{kumar_generalized_2025} by
decomposing gravity into components normal and tangential to the agar
surface. The surface-tangential component enters the lateral thin-film
transport as a downslope body-force contribution, whereas the
surface-normal component modifies the hydrostatic contribution to the
film pressure and thereby the pressure-dependent vertical exchange
between the agar and the swarm film
(Fig.~\ref{fig:model_schematic}). The complete governing equations and
numerical implementation are provided in the Supplementary Information.
All model quantities in this subsection are nondimensional unless otherwise stated. For readability, the superscript \(^{*}\) is retained only on the effective gravity couplings \(G_0^{*}\), \(G_n^{*}\), and \(G_t^{*}\) to distinguish them from the dimensional gravitational acceleration \(g\).

\begin{figure}[htbp]
\centering
\includegraphics[width=0.85\textwidth]{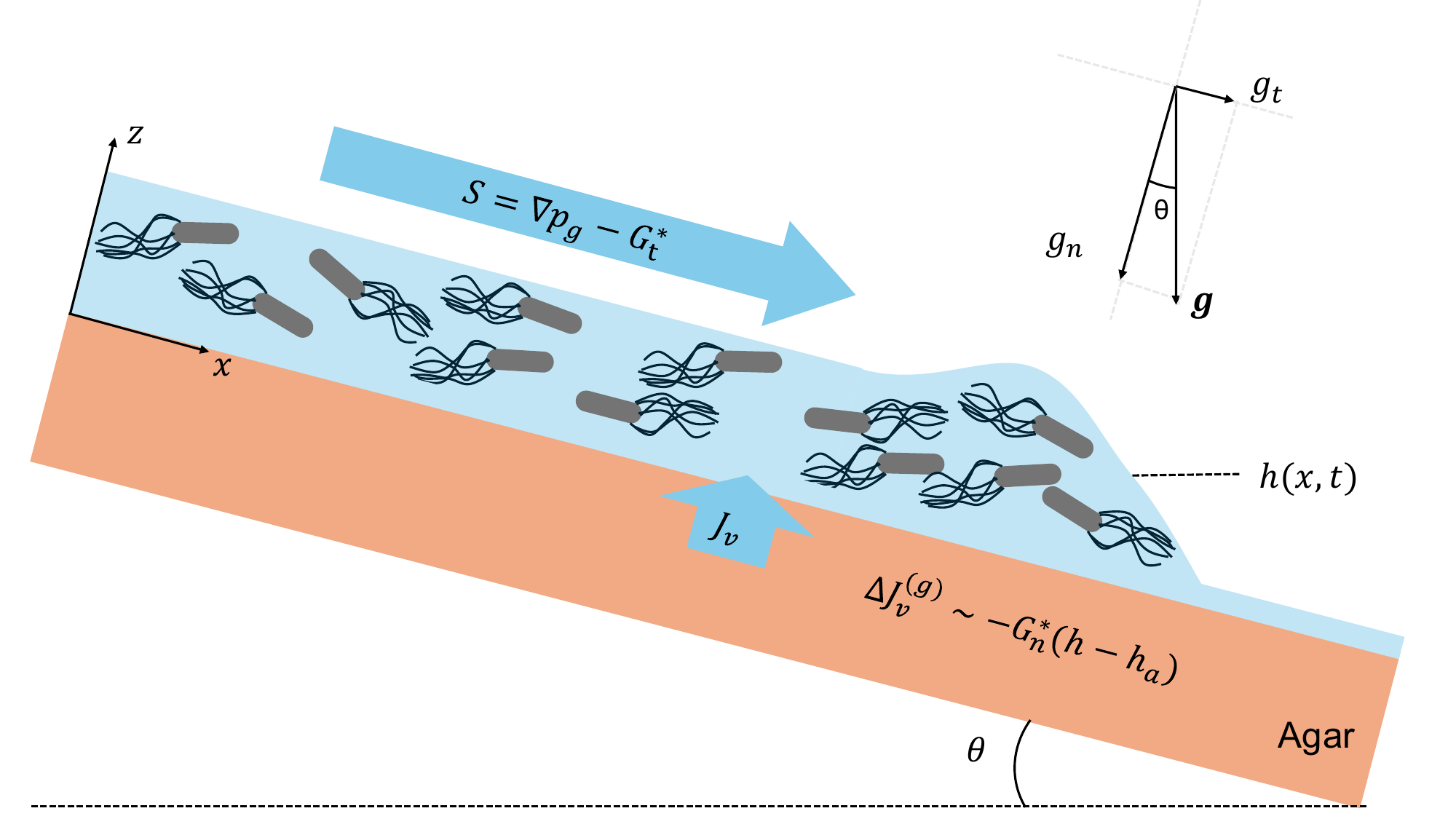}
\caption{
Schematic of the gravity-projected transport model on an inclined agar
surface. The diagram is a cross-sectional representation, with \(x\)
tangential and \(z\) normal to the physical agar surface, and
\(h(x,t)\) denoting the local film thickness. The numerical model
evolves \(h(x,y,t)\) over the agar-surface plane and does not explicitly
discretize the \(z\)-direction. The dimensional gravity \(g\) is
resolved into the physical projections \(g_t\) and \(g_n\). The
corresponding nondimensional model couplings are
\(G_t^{*}\) and \(G_n^{*}\), obtained by projecting
\(G_0^{*}\). The surface-tangential coupling enters
\(\mathbf S=\nabla p_g-G_t^{*}\), whereas the normal
coupling modifies \(p_g\) and the pressure-dependent exchange flux
\(J_v\). The schematic relation
\(\Delta J_{v,g}\propto-G_n^{*}(h-h_a)\) denotes the
gravity-dependent contribution to the pressure-feedback pathway.
Complete definitions are provided in the Supplementary Information.
}
\label{fig:model_schematic}
\end{figure}

For each inclination angle, the model was evolved to the same final
simulation time using a common parameter set. The effective
nondimensional gravity-coupling strength was fixed at
\(G_0^{*}=5.0\times10^{-6}\) for all inclination conditions. Its order of
magnitude was estimated from a gravity-to-capillarity scaling,
\(G_0^{*}\sim\rho gL^2/\gamma_{\mathrm{eff}}\), as detailed in the
Supplementary Information. Because the reduced model was not calibrated
to absolute experimental length or time scales, comparisons focus on
relative angular-response patterns rather than absolute dimensional
values.

To separate the roles of the two gravity projections, we also performed
two gravity-component isolation simulations using the same initial
condition, parameter set, diagnostic thresholds and final simulation
time. In the \(G_n^{*}\)-only simulation, the surface-tangential component
was removed by setting \(G_t^{*}=0\), while \(G_n^{*}\) was
retained in the gravity-dependent pressure and the pressure-dependent
vertical exchange pathway. In the
\(G_t^{*}\)-only simulation, the normal component was
removed by setting \(G_n^{*}=0\), while
\(G_t^{*}\) was retained in the lateral transport terms.

The normal and surface-tangential gravity contributions are defined as
\begin{subequations}
\label{eq:gravity_components}
\begin{align}
G_n^{*}
&=
G_0^{*}\cos\theta,
\\
G_t^{*}
&=
G_0^{*}\sin\theta\,\mathbf e_d,
\end{align}
\end{subequations}
where \(G_0^{*}\) is the fixed effective nondimensional gravity-coupling
strength, \(G_t^{*}\) denotes the surface-tangential vector coupling,
and \(\mathbf e_d\) is the unit vector directed downhill along the agar
surface. Thus, angle dependence is introduced only through
the geometric projections \(\cos\theta\) and \(\sin\theta\), rather
than through independently fitted angle-dependent forcing amplitudes.
\begin{figure}[htbp]
    \centering
    \includegraphics[width=0.95\textwidth]{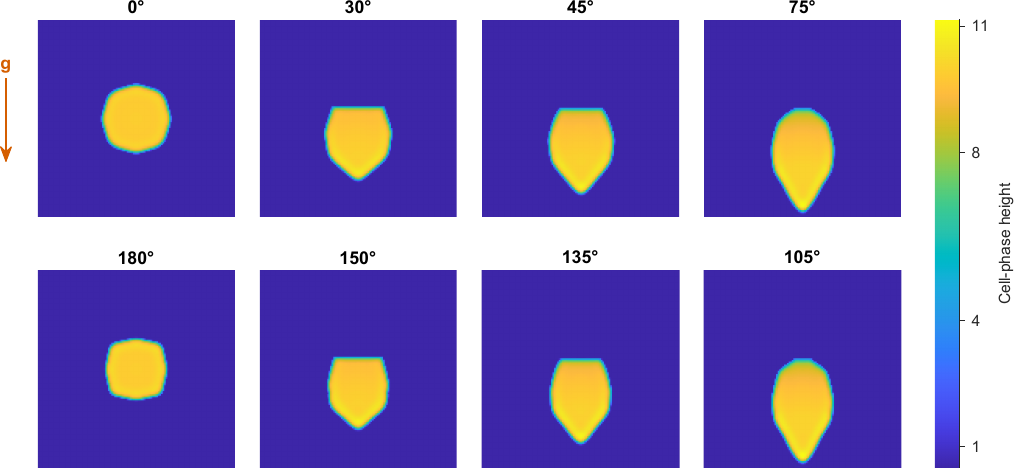}
\caption{
Final simulated cell-phase morphologies at a common nondimensional
simulation time. Color denotes the nondimensional cell-phase height
\(h_1\), with the same scale used for all inclination conditions.
Morphology was visualized from the cell-phase field, whereas the area
trajectories in Fig.~\ref{fig:sim_area_angle} were quantified from the
total wetting-front field. The field \(h_1\) represents local
cell-phase accumulation and is not a direct measurement of bacterial
number density. Angles are arranged in supplementary pairs, with
\(0^\circ\)--\(75^\circ\) in the upper row and the corresponding
\(180^\circ\)--\(105^\circ\) configurations in the lower row; gravity
acts downward in the displayed coordinate system.
}
    \label{fig:sim_final_morphology}
\end{figure}

\begin{figure}[tp]
\centering
\includegraphics[width=0.95\textwidth]{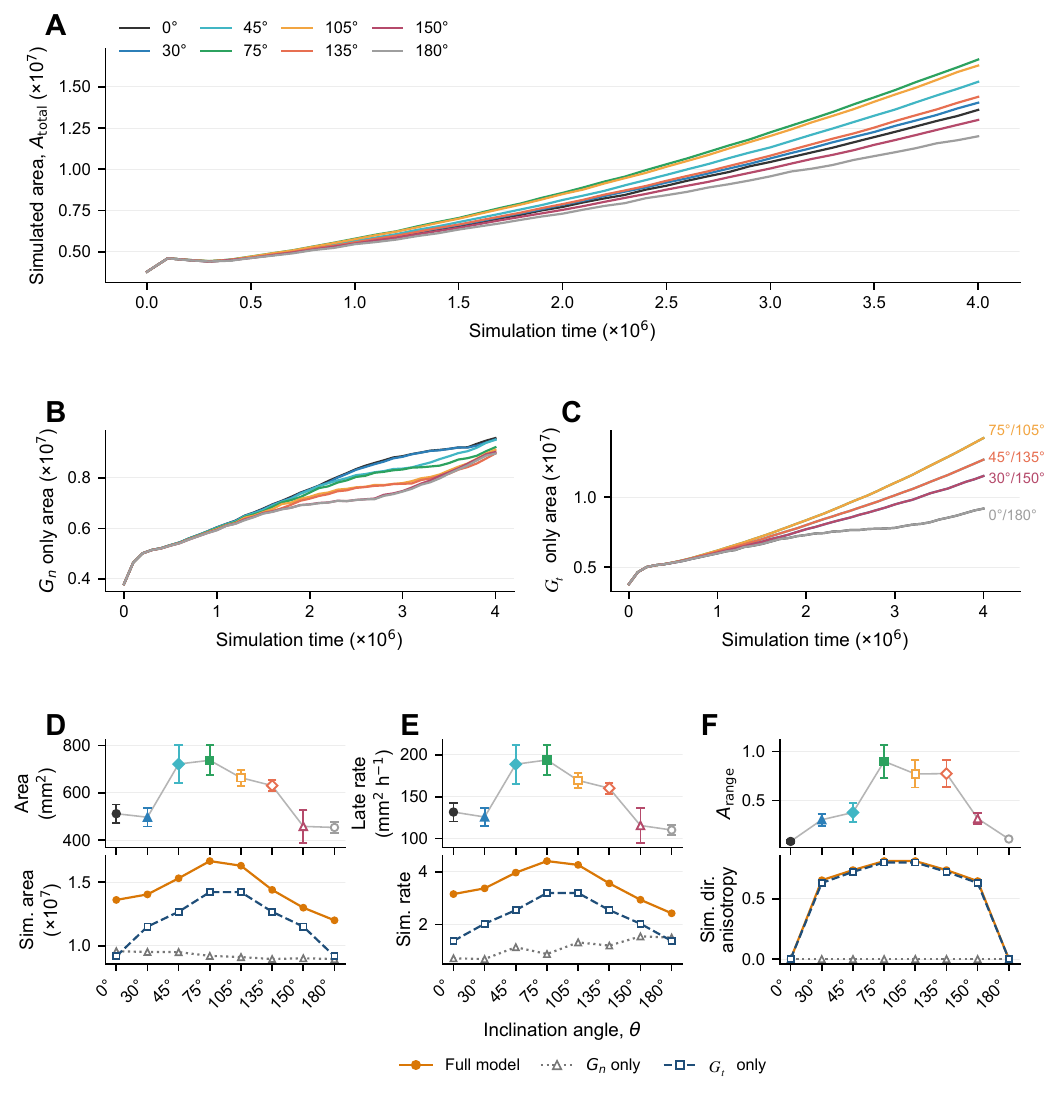}
\caption{
Full-model and gravity-component isolation simulations separate the
roles of surface-normal and surface-tangential gravity coupling in
wetting-front expansion and directional deformation.
\textbf{A}, Nondimensional wetting-front area
\(A_{\mathrm{total}}\) over nondimensional time \(t\) for the full
model.
\textbf{B}, Corresponding area evolution with only \(G_n^{*}\)
retained.
\textbf{C}, Area evolution with only
\(G_t^{*}\) retained; supplementary angle pairs
collapse because they have the same \(|\sin\theta|\).
\textbf{D}, Experimental final colony area and simulated final
wetting-front area.
\textbf{E}, Experimental late-stage area expansion rate and the
corresponding simulated late-growth measure.
\textbf{F}, Experimental range-over-mean anisotropy
\(A_{\mathrm{range}}\) and simulated \(A_{\mathrm{range,sim}}\).
Upper subpanels in \textbf{D--F} show experiments and lower subpanels
show simulations. Simulated time, area and growth rate are
nondimensional; the factors \(10^6\) and \(10^7\) are display scaling
factors only. The model was not fitted to absolute experimental length
or time scales, so comparison focuses on angular response patterns and
directional deformation.
}
\label{fig:sim_area_angle}
\end{figure}

The model qualitatively recapitulated the main structure of the experimental macroscopic response. First, simulated wetting-front area curves separated progressively over time, consistent with the experimental observation that inclination-dependent differences accumulated during expansion rather than appearing only at the final endpoint (Fig.~\ref{fig:sim_area_angle}A). Second, both experiment and simulation showed a non-monotonic dependence of expansion on inclination angle. In the experiments, the final area and the 6--9~h late expansion rate both reached their maximum at \(75^\circ\), and the simulation showed the same peak angle for both final wetting-front area and late growth rate (Fig.~\ref{fig:sim_area_angle}D,E).

The gravity-component isolation simulations clarified how the two gravity projections contributed to this response. The \(G_n^{*}\)-only model produced only weak separation of the area curves and little directional anisotropy (Fig.~\ref{fig:sim_area_angle}B,D--F). Thus, retaining only the surface-normal gravity contribution was insufficient to generate the strong directional deformation obtained in the full model. By contrast, the \(G_t^{*}\)-only model generated clear time-dependent area separation and strong directional anisotropy (Fig.~\ref{fig:sim_area_angle}C--F). Because
\(\sin\theta=\sin(180^\circ-\theta)\), supplementary angle pairs collapsed into the same responses. Therefore, in the model, the surface-tangential component is the main contributor to lateral redistribution and directional deformation, whereas the normal component breaks the supplementary-angle symmetry through the film-pressure terms and their coupling to water exchange.

This component-isolation result also explains why the full model distinguished paired up-facing acute and down-facing obtuse configurations. The \(G_t^{*}\)-only simulation contains the same lateral forcing magnitude for \(\theta\) and \(180^\circ-\theta\), and therefore tends to make each supplementary pair similar. In the full model, these paired configurations still share the same \(\lVert G_t^{*}\rVert\), but they differ in \(G_n^{*}=G_0^{*}\cos\theta\). The normal component changes the pressure state of the film and the vertical exchange term, allowing up-facing and down-facing configurations to produce different expansion responses.

The morphology comparison showed the same division of roles. The \(G_n^{*}\)-only simulation produced little directional anisotropy, whereas the \(G_t^{*}\)-only simulation generated strong tilted-state anisotropy with nearly symmetric supplementary pairs (Fig.~\ref{fig:sim_area_angle}F). The full model retained this tangentially driven directional deformation while allowing the normal component to modulate the angular response. Consistent with this, the final simulated morphologies were comparatively symmetric at \(0^\circ\) and \(180^\circ\), but became directionally deformed under inclined conditions (Fig.~\ref{fig:sim_final_morphology}). Thus, downhill-biased morphology in the model mainly arose from lateral gravity-driven transport, whereas the normal component adjusted the expansion strength through its effects on film pressure and pressure-dependent replenishment. 

We also examined whether inclination generated spatially heterogeneous local
states near the simulated colony edge. Sector-resolved sampling showed
variation in the nondimensional cell-phase height \(h_1\), aqueous-phase
height \(h_2\), and model cell-fraction proxy \(\phi\) around inclined
colonies (Supplementary Fig.~\ref{fig:si_local_states}). These fields
provide the model's local phase-state predictions.

The nondimensional radial fluxes of the cell and aqueous phases corresponded
to the experimental sector-resolved expansion rate
(Supplementary Fig.~\ref{fig:si_radial_transport}), with stronger
outward transport generally associated with faster local front
advancement within a given inclination. We use this comparison as a
descriptive spatial consistency check for a transport-based
interpretation. The modeled phase fluxes are coarse-grained transport
quantities and are not direct equivalents of the microscopic PIV
velocity field. Overall, the simulations support a substantial role for
gravity-biased transport in the macroscopic angular response while
remaining a simplified description of the full swarm.

\FloatBarrier

\section*{Discussion}

The macroscopic response is consistent with the coupled action of
surface-tangential transport and surface-normal pressure and water exchange.
In the model, the surface-tangential gravity contribution drives downhill
redistribution of the thin film, while the surface-normal contribution
modifies lateral pressure gradients and pressure-dependent exchange with the
porous agar. Because these contributions vary with \(\sin\theta\) and
\(\cos\theta\), supplementary acute and obtuse configurations experience
the same lateral forcing magnitude but different normal-pressure states.
This coupling provides a route to downhill-biased morphology and
acute--obtuse asymmetry. It also permits a non-monotonic area response as
tangential redistribution strengthens toward the intermediate near-vertical inclinations, while the
changing normal-pressure state and replenishment feedback influence the
fluid available for sustained spreading. Component-isolation simulations support
these roles within the model. This interpretation is consistent with studies
of water recruitment, osmotic influx, wettability and interfacial transport
in swarming \cite{wu_microbubbles_2011,wu_water_2012,ping_osmotic_2014,
seminara_osmotic_2012,trinschek_osmotic_2016,trinschek_front_2018,
srinivasan_multiphase_2019}. The model therefore provides a
coarse-grained transport explanation for the macroscopic angular response
without prescribing angle-specific motility parameters.

The microscopic activity response may reflect local conditions controlled
by hydration, cell packing, substrate friction and confinement. Thin or
poorly connected aqueous films can increase cell--substrate resistance,
while water and biomass accumulation may alter crowding and hydrodynamic
interactions \cite{dechesne_hydration_2010,tecon_film_2016,wu_water_2012,
ping_osmotic_2014,ilkanaiv_aspect_2017,beer_phase_2020}.
Dense regions may also face greater oxygen demand and longer diffusion
paths \cite{wessel_oxygen_2014,partridge_architecture_2018}. We hypothesize that inclination may shift
different peripheral regions away from favorable motion conditions through downhill
redistribution and pressure-dependent replenishment, contributing to
the generally reduced mean activity, its angular recovery and the
observed upper-to-tip speed decrease. The mechanism is coupled with too many factors to be fully understood here and out of scope of this paper.
Notably, expansion is strongest at some intermediate inclinations despite
reduced microscopic speed, and the downhill front advances faster where
the tip's mean motion is lower. Local velocity magnitude alone therefore
cannot account for front progression; directed cell and liquid supply and
wetting are additional plausible contributions
\cite{wu_water_2012,ping_osmotic_2014,beer_surfactant_growth_2009}.

The small signed tangential velocity indicates a weak edge-parallel bias,
more pronounced near the sampled edge and modulated by inclination and sector.
Related papers show that boundary-guided alignment and hydrodynamic
interactions can amplify weak directional tendencies
\cite{wu_microbubbles_2011,xu_selforganization_2019,li_robust_edge_2024}.
Inclination-driven redistribution of water and biomass could change local
packing, substrate coupling and alignment along the edge, thereby modulating
this bias. Its contribution in SM3 remains to be tested by measuring local
film conditions and cell orientation.

The mesoscopic measurements reveal a smaller characteristic flow scale near the downhill tip. Velocity correlations are shorter, while vortex cores are smaller and more numerous. Both correlation length and core size increase inward across the sampled regions. A possible origin of this spatial organization lies in the competition between downhill transport and local contact-line retention. Gravity-biased transport may continuously supply liquid and biomass to the downhill front, allowing a temporarily pinned region to accumulate material before undergoing localized release. Similar roles of fluid recruitment, wetting and interfacial deformation in bacterial depinning and invasion have been demonstrated in related systems
\cite{hennes_active_2017,xu_invasion_2023}. If such release occurs repeatedly, the accompanying changes in front geometry and fluid redistribution may continually disturb the adjacent collective flow. Developing vortices could then be displaced, disrupted or split before larger coherent structures emerge, resulting in shorter velocity correlations and a greater number of smaller vortex cores. Vortex attachment, splitting and disappearance, together with a shift toward smaller flow scales near an interface, have been observed in bacterial swarms
\cite{patteson_interface_2018}. This proposed competition between vortex organization and intermittent front release could explain how finer-scale flow near the downhill tip accompanies faster front advancement and occasional comet-like extensions. The mechanism remains hypothetical because individual depinning events and vortex lifetimes were not measured concurrently. Synchronized measurements of front displacement, film geometry and time-resolved PIV could test whether repeated release interrupts the development of larger vortices.

\section*{Conclusion}

Surface inclination altered the expansion and morphology of
\textit{Enterobacter} sp. SM3 swarms, producing a non-monotonic
expansion response and downhill-biased colony shapes. PIV showed
generally lower mean activity under inclination and inward increases in
speed and spatial velocity-correlation length, with slower motion, shorter correlations
and smaller vortex cores toward the downhill tip. A gravity-projected two-phase
thin-film model with pressure-dependent Darcy-type replenishment
qualitatively reproduced the main macroscopic trends. Together, these
results connect surface orientation with colony-scale transport and
the spatial organization of collective motion.

\section*{Materials and Methods}

\subsection*{Bacterial culture and swarming-plate preparation}

The bacterial strain used in this study was \textit{Enterobacter}
sp. SM3. Frozen stocks were streaked onto \(1.5\%\) LB agar before
each experiment. A single colony was used to inoculate liquid LB
medium, and overnight cultures were grown at
\(30~^\circ\mathrm{C}\) and 200~rpm for 13.5~h.

Swarming assays were performed on semi-solid LB agar containing
\(0.5\%\) agar. For each 90~mm dish, 9.5~mL of molten medium was
poured and allowed to solidify under identical conditions. Plates were
dried for 10~min in a laminar-flow hood and stored at
\(4~^\circ\mathrm{C}\) for 7 days before use to standardize agar
hydration and reduce premature liquid runoff during inclined
incubation.

A \(1.85~\mu\mathrm{L}\) droplet of SM3 culture was inoculated at the
center of each plate. Plates were first incubated horizontally for
3~h to establish an initial swarm and were then transferred to the
inclination setup.

\subsection*{Inclination assay}

Plates were maintained at one of eight agar-surface orientations
relative to gravity:
\[
0^\circ,\ 30^\circ,\ 45^\circ,\ 75^\circ,\ 105^\circ,\
135^\circ,\ 150^\circ,\ 180^\circ.
\]
The \(0^\circ\) and \(180^\circ\) conditions had no gravitational
component parallel to the agar surface and were used as flat reference
configurations. Acute angles corresponded to up-facing tilted surfaces,
whereas supplementary obtuse angles corresponded to down-facing tilted
surfaces.

For inclined plates, the projection of gravity onto the agar surface
defined the downhill direction. In the image coordinate system used
for directional analysis, downhill was assigned to
\(-\pi/2\) and uphill to \(\pi/2\). Plates were incubated at
\(30~^\circ\mathrm{C}\) and \(90\%\) relative humidity throughout the
inclination phase.

\subsection*{Macroscopic imaging and expansion metrics}

Macroscopic colony images were acquired once per hour from 3 to 9~h
after inoculation. Colony outlines were segmented in ImageJ and
processed using custom analysis scripts, with manual correction when
required. From each outline, we extracted colony area \(A(t)\) and
the directional radius \(R(\alpha,t)\), measured from the colony
center at polar angle \(\alpha\).

The area expansion rate over an interval \([t_1,t_2]\) was
\begin{equation}
v_A(t_1,t_2)
=
\frac{
A(t_2)-A(t_1)
}{
t_2-t_1
}.
\label{eq:area_rate}
\end{equation}
The late-stage area expansion rate reported in the main text was
\begin{equation}
v_A^{6-9}
=
\frac{
A(9~\mathrm{h})-A(6~\mathrm{h})
}{
3~\mathrm{h}
}.
\label{eq:late_area_rate}
\end{equation}

The directional radial expansion rate was
\begin{equation}
v_R(\alpha;t_1,t_2)
=
\frac{
R(\alpha,t_2)-R(\alpha,t_1)
}{
t_2-t_1
}.
\label{eq:radial_rate}
\end{equation}
Unless otherwise stated, directional rates were calculated over the
6--9~h interval and are denoted by \(v_R^{6-9}(\alpha)\).

\subsection*{Macroscopic shape metrics}

Overall colony-shape anisotropy was quantified using the
range-over-mean index
\begin{equation}
A_{\mathrm{range}}
=
\frac{
R_{\max}-R_{\min}
}{
\langle R\rangle
},
\label{eq:anisotropy}
\end{equation}
where \(R_{\max}\), \(R_{\min}\), and \(\langle R\rangle\) are the
maximum, minimum, and mean directional radii at the same time point.

Downhill-biased extension was quantified as
\begin{equation}
B_R
=
\frac{
\left\langle
R(-3\pi/4),R(-\pi/2),R(-\pi/4)
\right\rangle
}{
\left\langle
R(\pi/4),R(\pi/2),R(3\pi/4)
\right\rangle
}.
\label{eq:downhill_bias}
\end{equation}
Here, the numerator and denominator are the mean radii in the three
downhill-side and three uphill-side directions, respectively.
\(B_R=1\) indicates equal upper- and lower-side extension, whereas
\(B_R>1\) indicates preferential downhill extension.

Localization of late-stage expansion at the downhill tip was measured
using
\begin{equation}
C_{\mathrm{tip}}
=
\frac{
v_R^{6-9}(-\pi/2)
}{
\left\langle
v_R^{6-9}(-3\pi/4),
v_R^{6-9}(-\pi/4)
\right\rangle
}.
\label{eq:tip_localization}
\end{equation}
Values greater than one indicate faster expansion at the downhill tip
than at its two adjacent flanks.

\subsection*{Microscopic imaging and PIV processing}

Microscopic recordings were acquired with a Nikon Ti2-U microscope,
a \(60\times\) objective and a PCO Edge 5.5 sCMOS camera, using
\(1500\times1500\)~px windows. Stored time calibration corresponds
to 59 frames per second for most recordings. Spatial and velocity calibration and
sequence length were retained per recording. Filtered PIV fields were
obtained using PIVlab \cite{thielicke_pivlab_2021}. Individual cells
were not resolved in the dense swarm. The spatial calibration was
approximately \(0.10833~\mu\mathrm m\) per pixel, and vector spacing
was 16~px (\(1.73~\mu\mathrm m\)).

Recordings were taken at upper, upper-diagonal, side, lower-diagonal
and leading-tip sectors, with the imaged outer side oriented to the
right. The inward coordinate was measured from half a vector spacing
beyond the rightmost PIV column. The principal comparison interval was
100-1300~px, excluding boundary strips with restricted field coverage
or correlation neighborhoods. The 100-400~px and 1000--1300~px intervals
were used to localize outer-inner differences. For strip-based
quantities, 100-px strips were weighted by their actual native-column
counts when computing interval means. 

\subsection*{Velocity measures and spatial organization}

Mean speed was calculated from valid PIV-vector magnitudes. RMS vorticity
was obtained from in-plane velocity gradients as a supporting activity
measure. In the adopted image coordinates, \(v_n=v_x\) and \(v_t=v_y\),
with positive \(v_n\) pointing toward the outer side. The sign of \(v_t\)
was not independently calibrated to clockwise or counterclockwise motion
in the colony plane. Signed averages were formed within each recording
before biological aggregation, as defined in the Supplementary Information.

The spatial velocity-correlation length, \(L_{vv}^{\parallel}\), uses
dot products of the two-component velocity at edge-parallel separations
within depth strips. The initial positive correlation lobe was integrated
without subtracting the mean velocity. Valid-pair weighting and the
handling of estimates without a resolved zero crossing are defined in
the Supplementary Information.

Vortices were identified through instantaneous rotating cores from
\(\Gamma_1\) centers and \(\Gamma_2\) regions
\cite{graftieaux_combining_2001} in 16 frames per recording. Equivalent
core diameter and the number of detected vortices per eligible area were
averaged within recordings before plate-sector aggregation. Diameter
refers to the detected core, not a full closed-streamline envelope.
Preprocessing, multiscale detection and boundary exclusions are detailed
in the Supplementary Information.

\subsection*{Biological replication and statistical analysis}

Fifteen independent plates were analyzed per inclination. Repeated recordings were averaged within each plate and sector, and plate-level summaries gave equal weight to the available sectors. Eight-angle comparisons used Kruskal-Wallis tests. Sector and inclination-sector effects under tilted conditions were tested using models with plate-cluster-robust covariance. 
\section*{Acknowledgements}
We thank Professor Manoj M. Varma (Indian Institute of Science) for sharing Harshitha S. K.'s doctoral thesis, whose observations of bacterial depinning on inclined surfaces helped our modeling and discussion of downhill swarm expansion. Our work was supported by the National Natural Science Foundation of China (NSFC), Grant No. 12202275.


\section*{Competing interests}
The authors declare no competing interests.



\clearpage
\nolinenumbers
\setcounter{page}{1}
\setcounter{figure}{0}
\setcounter{equation}{0}
\setcounter{table}{0}
\setcounter{footnote}{0}
\setcounter{section}{0}
\setcounter{subsection}{0}
\setcounter{subsubsection}{0}
\renewcommand{\thefigure}{S\arabic{figure}}
\renewcommand{\theequation}{S\arabic{equation}}
\renewcommand{\thetable}{S\arabic{table}}
\renewcommand{\topfraction}{0.7}
\renewcommand{\bottomfraction}{0.3}
\renewcommand{\textfraction}{0.2}
\renewcommand{\floatpagefraction}{0.5}
\setlength{\textfloatsep}{20pt plus 2pt minus 4pt}
\setlength{\intextsep}{12pt plus 2pt minus 2pt}
\setlength{\floatsep}{12pt plus 2pt minus 2pt}
\captionsetup{font=small,labelfont=bf,skip=10pt}

\begin{center}
{\LARGE\bfseries Supplementary Information\\[0.5em]
Inclined surfaces bias bacterial swarming expansion and edge organization\par}

\vspace{1em}

Pengxi Gu, Zixiang Lin, Hui Ren, Jin Zhu, Anqi Li, Omar Said, Weijie Chen and Zijie Qu
\end{center}

\section*{Supplementary Information}


\subsection*{Imaging geometry and biological aggregation}

Microscopic recordings sampled the upper, upper-diagonal, side,
lower-diagonal and leading-tip regions. The two flat configurations
served as nondirectional references. With the swarm edge on the right,
the inward coordinate was
\begin{equation}
d=x_{\max}+\tfrac12\Delta x-x,
\label{eq:si_fullfield_depth}
\end{equation}
where \(x_{\max}\) is the rightmost PIV-column position and \(\Delta x\)
is the native vector spacing. Spatial calibration was
\(0.10833~\mu\mathrm m\) per pixel, and vector spacing was
\(1.73~\mu\mathrm m\). The edge origin was fixed by the PIV grid, approximately corresponding to the swarm edge contact line. The common comparison interval
was 10.8--140.8~\(\mu\mathrm m\) (100--1300 px). Boundary strips
outside this interval were excluded because of restricted neighborhood
support or field coverage. The outer and inner intervals for signed
\(v_t\) were 10.8--43.3 and 108.3--140.8~\(\mu\mathrm m\),
respectively.

Fifteen independently prepared plates were analyzed per inclination.
Repeated recordings were averaged within plate-sector, and angle-level
summaries assigned equal weight to each observed sector of a plate.
Unavailable sectors were not imputed. Vectors, frames and detected
vortex cores were nested measurements, not independent replicates.
Error bars and bands show s.e.m. across plates unless stated otherwise.
Speed and spatial-correlation profiles were calculated in consecutive
10.8-\(\mu\mathrm m\) (100-px) strips. 

\subsection*{Mean speed, signed tangential velocity and vorticity}

Mean speed was the mean valid-vector magnitude. In the image coordinates,
\(v_n=v_x\) and \(v_t=v_y\), with positive \(v_n\) pointing toward
the outer side. The normal component is averaged over the same validated
10.8--140.8~\(\mu\mathrm m\) interval as the other sampled-region measures;
it does not represent the front-propagation velocity. For each recording \(r\) and native PIV column \(j\),
\begin{equation}
\mu_{rj}=\langle v_t\rangle_{rj},
\label{eq:si_native_mean}
\end{equation}
where the average uses accepted spatial samples and frames in that
column. The signed tangential means plotted in the main text average \(\mu_{rj}\)
over columns in the specified interval, then recordings and biological
units. The plate-level means in the outer and inner intervals were
\(-0.297\pm0.037\) and \(-0.102\pm0.032~\mu\mathrm m\,\mathrm s^{-1}\),
respectively. These interval means use the same biological aggregation
as the common-window comparisons.

In-plane vorticity and its frame-wise RMS in strip \(b\) were
\begin{equation}
\omega=\partial_xv_y-\partial_yv_x,\qquad
\Omega_{\mathrm{RMS}}(b,t)=\sqrt{\langle\omega^2\rangle_{b,t}}.
\label{eq:si_vorticity_rms}
\end{equation}
Finite differences were evaluated on the calibrated grid. Only for
derivative evaluation, invalid velocities were linearly filled along
rows and columns with nearest-value endpoint extension; any remaining
non-finite values were set to zero. Derivatives were retained only at
originally valid vectors with at least six valid cells in the surrounding
\(3\times3\) neighborhood. RMS values were averaged over frames,
then strips with native-column-count weights, and finally recordings
and plates. Figure~\ref{fig:si_activity_support}C,D shows the resulting
supporting activity measurement.

\subsection*{Spatial velocity correlations}

Within each depth strip, valid space--time pairs separated along
\(\hat{\mathbf t}\), the local edge tangent, give
\begin{equation}
C_{vv}^{\parallel}(r;d)=
\frac{\langle\mathbf v(\mathbf x,t)\cdot
\mathbf v(\mathbf x+r\hat{\mathbf t},t)\rangle_{d,r}}
{\langle|\mathbf v|^2\rangle_{d,0}},
\label{eq:si_fullfield_parallel_correlations}
\end{equation}
where \(\mathbf v=(v_n,v_t)\). Pair products and eligible-pair counts
were accumulated across columns and frames before normalization at
zero lag, using a common valid mask for both components. At least 16
pairs were required at each retained lag, which was limited to
\(65.0~\mu\mathrm m\) (600 px). The mean velocity was not subtracted.
The integral length was
\begin{equation}
L_{vv}^{\parallel}(d)=\int_0^{r_{0,vv}}
C_{vv}^{\parallel}(r;d)\,\mathrm dr.
\label{eq:si_fullfield_parallel_lengths}
\end{equation}
The first zero was linearly interpolated. The length measures correlations at along-edge separations,
with position \(d\) varying inward.

\subsection*{Vortex identification and core measurements}

The detector used the \(\Gamma_1\)--\(\Gamma_2\) approach
\cite{SIREF-graftieaux_combining_2001}. For point \(P\), neighboring point
\(M\), displacement \(\mathbf r_{PM}\) and neighborhood \(S(P)\),
\begin{align}
\Gamma_1(P)&=\frac1N\sum_{M\in S(P)}
\frac{[\mathbf r_{PM}\times\mathbf v(M)]_z}
{|\mathbf r_{PM}|\,|\mathbf v(M)|},\\
\Gamma_2(P)&=\frac1N\sum_{M\in S(P)}
\frac{[\mathbf r_{PM}\times(\mathbf v(M)-\overline{\mathbf v}_{S(P)})]_z}
{|\mathbf r_{PM}|\,|\mathbf v(M)-\overline{\mathbf v}_{S(P)}|}.
\label{eq:si_fullfield_gamma}
\end{align}
Here \(N\) counts eligible neighbors with nonzero denominators,
\(\overline{\mathbf v}_{S(P)}\) is the neighborhood mean velocity,
and the \(z\) subscript denotes the component normal to the image
plane. Each velocity component was first smoothed with the
mask-normalized kernel
\begin{equation}
K=\frac1{16}\begin{pmatrix}1&2&1\\2&4&2\\1&2&1\end{pmatrix}.
\label{eq:si_fullfield_core_smoothing}
\end{equation}
Disk radii were 3, 5 and 7 vector cells. Candidate centers were local
maxima of \(|\Gamma_1|>0.75\). Core boundaries used
\(|\Gamma_2|\ge2/\pi\), with the same sign as the candidate center.
Eight-connected components with fewer than six vector cells were
excluded. Each candidate also satisfied the rotation-dominance check
\begin{equation}
Q/\omega^2>0.02,\qquad Q=\tfrac12(\omega^2-s^2),\qquad
s^2=2\sum_{i,j}S_{ij}^2,\quad
S_{ij}=\tfrac12(\partial_i v_j+\partial_j v_i).
\label{eq:si_fullfield_rotation_check}
\end{equation}

Detections from different radii were ranked by center strength times
the square root of core area. Same-sign candidates were merged when
their centers were too close or their core-mask intersection-over-union
exceeded 0.20. The minimum center separation was the larger of two
vector spacings and \(0.30(D_i+D_j)\). Sixteen uniformly distributed
frames were analyzed per recording. Detected objects were assigned
to depth strips by center position. For core area \(A_c\),
\begin{equation}
D_c=2\sqrt{A_c/\pi}.
\label{eq:si_fullfield_core_diameter}
\end{equation}
Boundary-truncated cores were excluded from size statistics. Density
was the number of assigned centers divided by eligible detection area,
reported per \(1000~\mu\mathrm m^2\). Metrics were averaged over
frames and recordings before plate-sector aggregation. 

\subsection*{Regional gradients and statistical comparisons}

\begin{figure}[p]
\centering
\includegraphics[width=.97\textwidth]{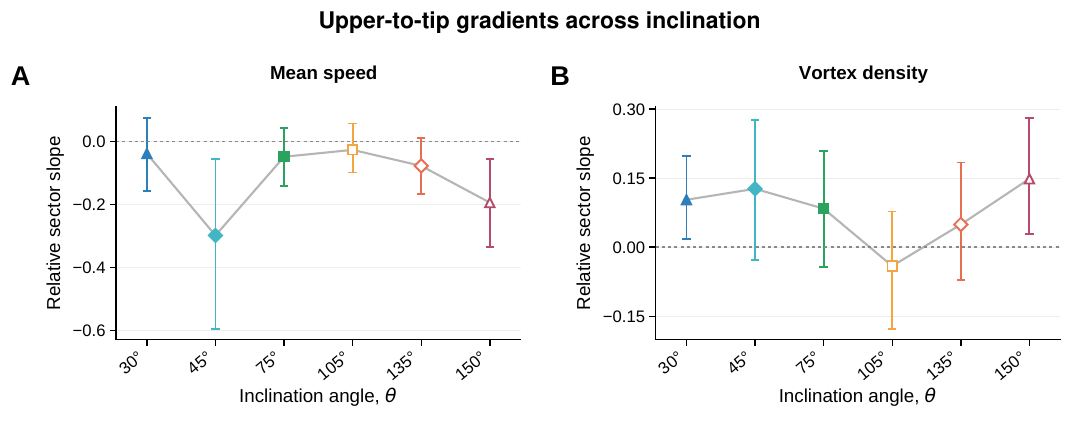}
\caption{Inclination-dependent summaries of upper-to-tip gradients.
\textbf{A}, Mean speed. \textbf{B}, Vortex density. Each slope is
estimated from a plate's observed ordered sector positions and
normalized by its sector mean. Points show mean plate slopes;
error bars are 95\% percentile intervals from 10,000 plate bootstrap
resamples. These descriptive slopes summarize the regional patterns;
the corresponding categorical inclination--sector tests appear in
Table~\ref{tab:si_cluster_tests}.}
\label{fig:si_regional_support}
\end{figure}

For the regional gradients in Fig.~\ref{fig:si_regional_support},
sector coordinates were \(q_s=0,1/4,1/2,3/4,1\), ordered from upper
to tip. Ordinary least squares on each plate's observed sectors gave
slope \(\beta\). The displayed slope was \(g=\beta/M\), where
\(M\) is the plate's sector mean. Intervals are 95\% percentile
intervals from 10,000 plate-level bootstrap resamples. Differences
between displayed slope means are descriptive; the categorical
inclination--sector tests are reported separately.

Eight-angle comparisons used Kruskal--Wallis tests. For the six tilted
conditions, categorical angle and sector models used
plate-cluster-robust covariance, with separate additive angle, additive
sector and interaction tests. BH-adjusted probabilities retain the
complete ten-metric family for eight-angle tests and the separate
nine-metric families for each model effect. Tables~\ref{tab:si_angle_tests}
and \ref{tab:si_cluster_tests} display only the five measures used in
the main-text argument. 

\begin{table}[htbp]
\centering
\small
\caption{Eight-angle tests for the reported microscopic measures.
Kruskal--Wallis tests have seven degrees of freedom. All measures
summarize 10.8--140.8~\(\mu\mathrm m\); the BH probabilities are
unchanged from the complete ten-metric family in the source data.}
\label{tab:si_angle_tests}
\begin{tabular}{lrrr}
\toprule
Metric & \(H_7\) & Raw \(p\) & BH-adjusted \(p\)\\
\midrule
Mean speed & 25.69 & 0.0006 & 0.0029\\
Signed \(v_t\) & 17.76 & 0.0131 & 0.0437\\
Spatial \(L_{vv}^{\parallel}\) & 6.42 & 0.4912 & 0.5696\\
Vortex-core diameter & 3.14 & 0.8720 & 0.8720\\
Vortex density & 6.92 & 0.4374 & 0.5696\\
\bottomrule
\end{tabular}
\end{table}

\begin{table}[H]
\centering
\small
\caption{Plate-cluster-robust tests in the six tilted conditions.
Entries are BH-adjusted probabilities retained from separate
nine-metric families for each effect. Wald statistics have 5, 4 and
20 degrees of freedom for additive angle, additive sector and
interaction, respectively. Full statistics are supplied in the source data.}
\label{tab:si_cluster_tests}
\begin{tabular}{lrrr}
\toprule
Metric & Angle, additive & Sector, additive & Interaction\\
\midrule
Mean speed & 0.0010 & \(1.76\times10^{-9}\) & 0.1908\\
Signed \(v_t\) & 0.0135 & 0.0060 & 0.0961\\
Spatial \(L_{vv}^{\parallel}\) & 0.5193 & \(5.52\times10^{-6}\) & 0.5119\\
Vortex-core diameter & 0.4059 & \(4.29\times10^{-13}\) & 0.3458\\
Vortex density & 0.5292 & \(8.55\times10^{-8}\) & 0.0111\\
\bottomrule
\end{tabular}
\end{table}
\begin{figure}[H]
\centering
\includegraphics[width=.97\textwidth]{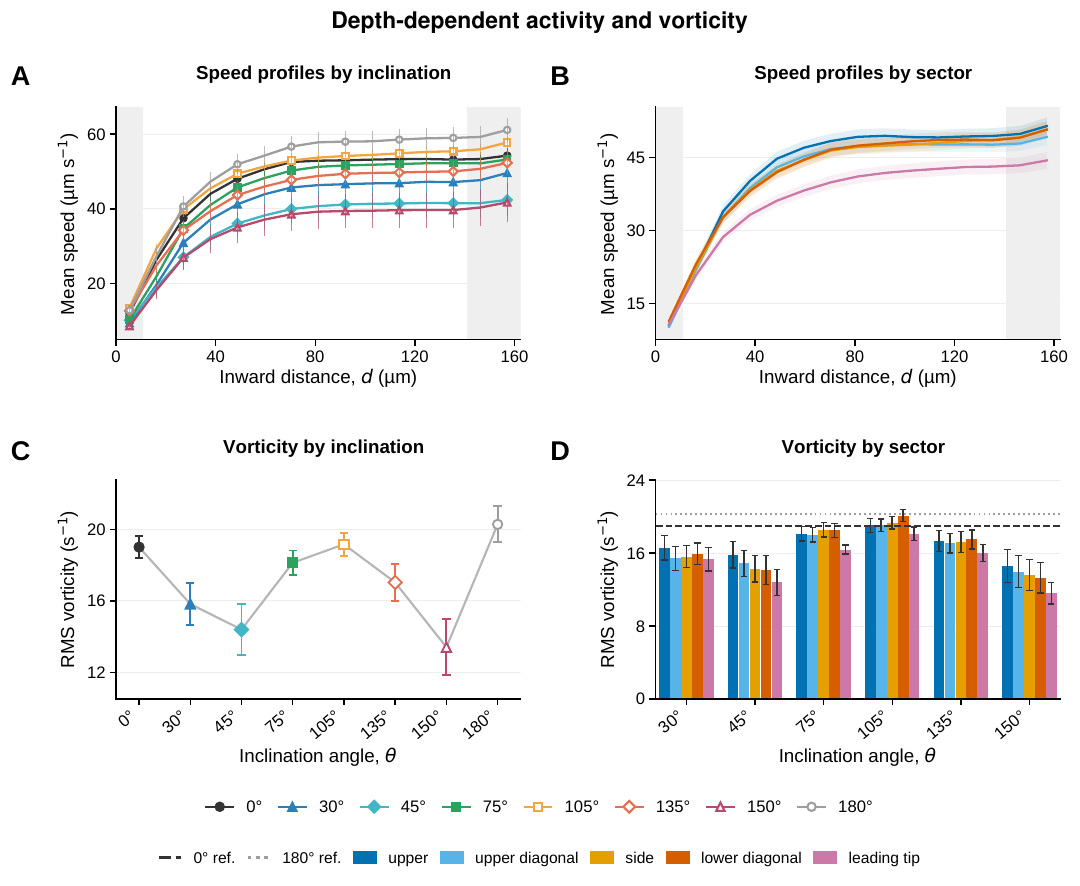}
\caption{Depth-dependent speed and supporting vorticity measurements.
\textbf{A,B}, Mean-speed profiles grouped by inclination and sector.
\textbf{C,D}, RMS vorticity averaged over 10.8--140.8~\(\mu\mathrm m\),
by inclination and sector. Inward distance increases from the swarm
edge on the right-hand side of the recorded field. Gray boundary strips
are outside the common comparison interval. Curves, bars and errors
show plate-level mean \(\pm\) s.e.m.; dashed and dotted horizontal
lines in D denote the flat-reference means. Sector profiles in B
summarize tilted conditions.}
\label{fig:si_activity_support}
\end{figure}
\clearpage

\clearpage

\begin{figure}[p]
\centering
\includegraphics[width=.97\textwidth,height=.81\textheight,keepaspectratio]{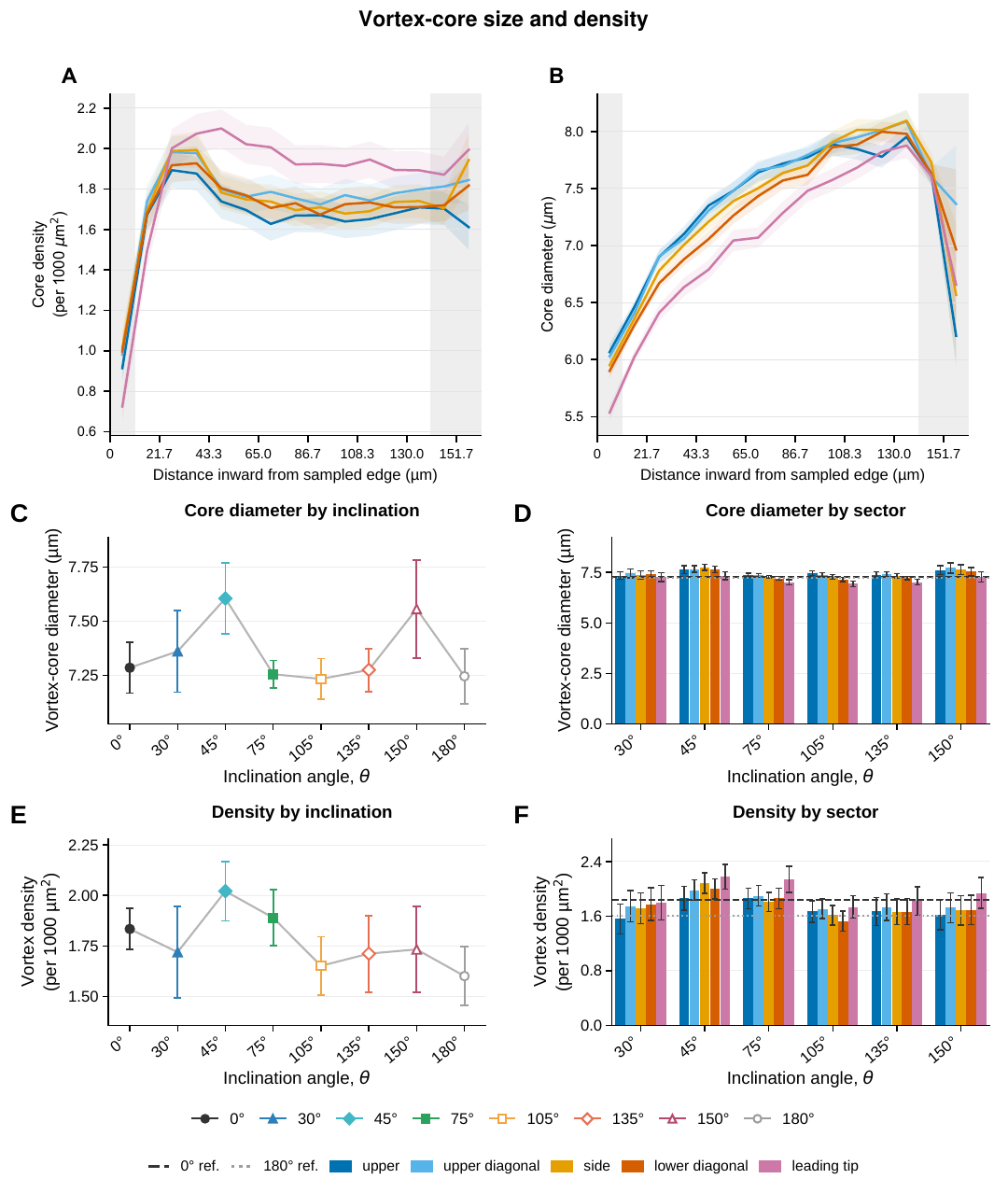}
\caption{Vortex-core size and density across depth, inclination and sector.
\textbf{A,B}, Sector-resolved depth profiles of detected-vortex density
and equivalent core diameter. \textbf{C,D}, Core diameter over
10.8--140.8~\(\mu\mathrm m\), by inclination and sector.
\textbf{E,F}, Vortex density over the same interval. Density is reported
per \(1000~\mu\mathrm m^2\) of eligible detection area. Curves, bars
and error bars or bands show plate-level mean \(\pm\) s.e.m.; gray strips
are excluded from the common interval. Colors and symbols follow the
main figures, and dashed and dotted reference lines denote the flat
conditions. Individual cores are not biological replicates.}
\label{fig:si_vortex_support}
\end{figure}
\clearpage

\begin{figure}[p]
\centering
\includegraphics[width=.97\textwidth]{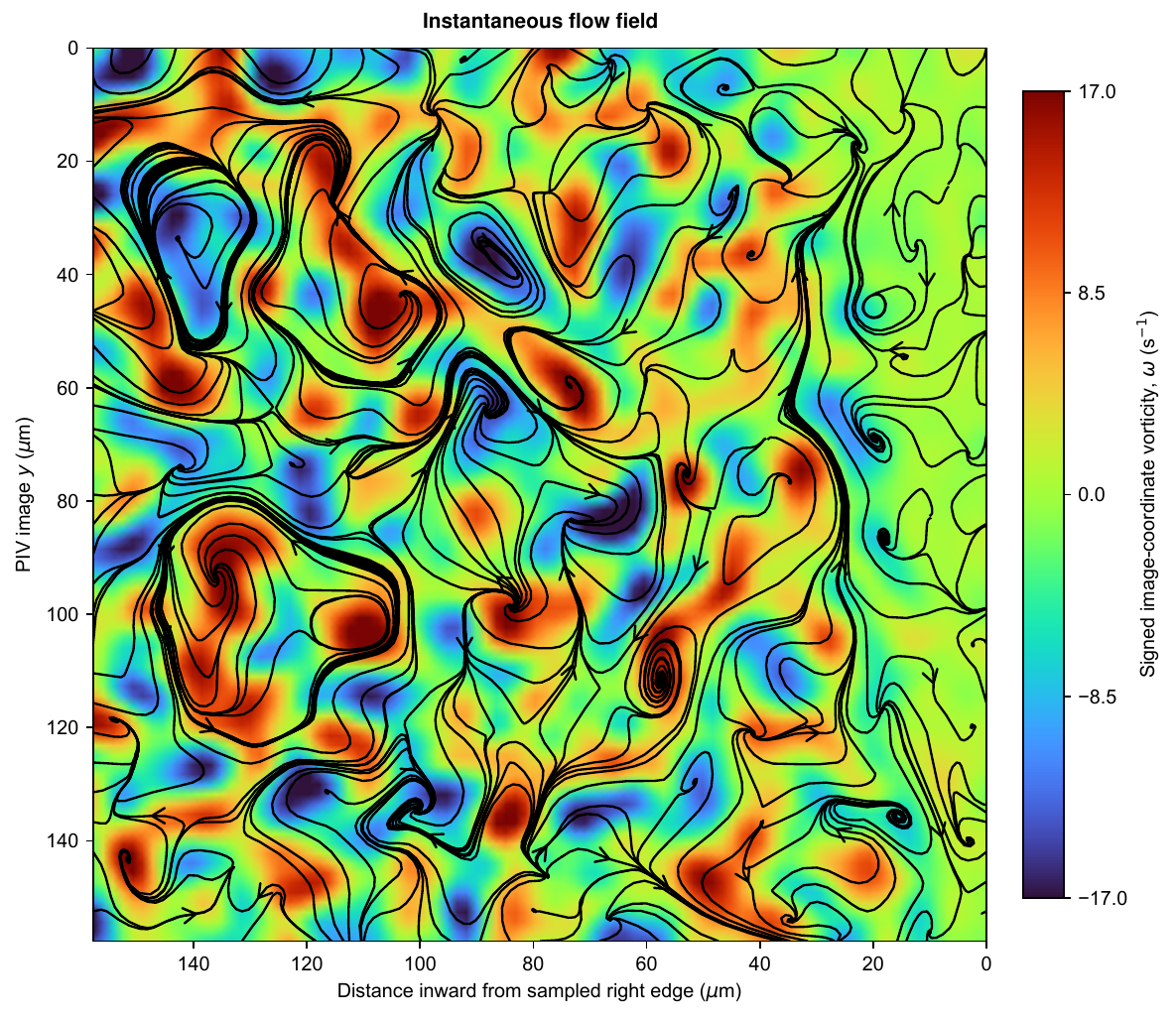}
\caption{Illustrative instantaneous flow field. Black streamlines are
shown over image-coordinate vorticity, with a color range of
\(-17\) to \(17~\mathrm s^{-1}\). The swarm edge is on the right
and inward distance increases to the left; image \(y\) increases
downward. The image illustrates instantaneous flow organization.
The reported mean tangential velocities are quantified separately
from complete recordings.}
\label{fig:si_streamline_example}
\end{figure}
\clearpage

\subsection*{Modeling and simulation}

\subsubsection*{Nondimensional notation and model overview}

All variables and parameters in this section are nondimensional unless
explicitly identified as dimensional. For readability, nondimensional
markers are omitted from the model coordinates, fields, pressures,
fluxes, lengths, times and diagnostic quantities after this statement.
The superscript \(^{*}\) is retained only on the effective gravity
couplings \(G_0^{*}\), \(G_n^{*}\), and
\(G_t^{*}\) to distinguish them from the dimensional
gravitational acceleration \(g\) used in the scaling estimate.
Numerically regularized heights are denoted by the subscript
``reg.''

Let \(x\) and \(y\) denote Cartesian coordinates in the agar-surface
plane, \(t\) denote simulation time, and \(\mathcal D\) denote the
circular computational domain. We model
the swarm as a thin two-phase film composed of a cell phase and an
aqueous phase. The baseline structure follows Kumar and
Pushpavanam~\cite{SIREF-kumar_generalized_2025} and includes cell growth,
aqueous influx, substrate wettability, surfactant production and
Marangoni-driven transport. Inclination is introduced through one
nondimensional gravity-coupling strength \(G_0^{*}\), projected normal
and parallel to the agar surface.

\subsubsection*{State variables and numerical regularization}

The cell-phase height, aqueous-phase height, total film height and
interfacial surfactant concentration are
\begin{equation}
h_1(x,y,t),
\qquad
h_2(x,y,t),
\qquad
h=h_1+h_2,
\qquad
\Gamma(x,y,t).
\label{eq:si_state_variables}
\end{equation}
An auxiliary nondimensional substrate-potential field
\(\psi(x,y,t)\) is also retained in the general solver.
The local cell-fraction proxy is
\begin{equation}
\phi
=
\frac{h_{1,\mathrm{reg}}}
{h_{1,\mathrm{reg}}+h_{2,\mathrm{reg}}},
\label{eq:si_cell_fraction}
\end{equation}
where
\begin{equation}
h_{\mathrm{reg}}
=
\max(h,h_{\min}),
\qquad
h_{1,\mathrm{reg}}
=
\max(h_1,h_{1,\min}),
\qquad
h_{2,\mathrm{reg}}
=
\max(h_2,h_{2,\min}).
\label{eq:si_regularized_heights}
\end{equation}
These floors prevent singular denominators and negative-power
divergences in precursor regions and contain no angle dependence.

\subsubsection*{Two-phase thin-film transport}

The baseline nondimensional film pressure is
\begin{equation}
p_0
=
-\nabla^2h
+
W\left[
(h_{\mathrm{reg}})^{-3}
-
(h_{\mathrm{reg}})^{-6}
\right],
\label{eq:si_baseline_pressure}
\end{equation}
where \(W\) controls the disjoining-pressure contribution. On a
horizontal substrate, the pressure-driving field is
\begin{equation}
\mathbf S_0=\nabla p_0.
\label{eq:si_S0}
\end{equation}

The cell, aqueous and surfactant equations are
\begin{subequations}
\label{eq:si_transport_system}
\begin{align}
\frac{\partial h_1}{\partial t}
&=
\nabla\!\cdot
\left[
M_{1p}\mathbf S
+
M_{1\Gamma}\nabla \Gamma
\right]
+
R_c,
\\
\frac{\partial h_2}{\partial t}
&=
\nabla\!\cdot
\left[
M_{2p}\mathbf S
+
M_{2\Gamma}\nabla \Gamma
\right]
+
J_v,
\\
\frac{\partial\Gamma}{\partial t}
&=
\nabla\!\cdot
\left[
M_{5p}\mathbf S
+
\left(M_{5\Gamma}+D\right)
\nabla \Gamma
\right]
+
R_{\Gamma}.
\end{align}
\end{subequations}
The mobilities are
\begin{subequations}
\label{eq:si_mobilities}
\begin{align}
M_{1p}
&=
\frac{h_{1,\mathrm{reg}}}{h_{2,\mathrm{reg}}}
\frac{(h_{\mathrm{reg}})^3}{3}
+
\beta h_{1,\mathrm{reg}},
&
M_{1\Gamma}
&=
\frac{h_{1,\mathrm{reg}}}{h_{2,\mathrm{reg}}}
\frac{(h_{\mathrm{reg}})^2}{2},
\\
M_{2p}
&=
\frac{(h_{\mathrm{reg}})^3}{3},
&
M_{2\Gamma}
&=
\frac{(h_{\mathrm{reg}})^2}{2},
\\
M_{5p}
&=
\Gamma
\left[
\frac{(h_{\mathrm{reg}})^3}
{2h_{2,\mathrm{reg}}}
+
\beta\frac{h_{1,\mathrm{reg}}}{h_{\mathrm{reg}}}
\right],
&
M_{5\Gamma}
&=
\Gamma
\frac{(h_{\mathrm{reg}})^2}
{h_{2,\mathrm{reg}}}.
\end{align}
\end{subequations}
Here \(D\) is the surfactant diffusion coefficient and \(\beta\) is
the additional cell-associated mobility coefficient.

Cell growth is
\begin{equation}
R_c
=
k_c h_1
\left(1-\frac{h_1}{p_{20}}\right)
f_{\mathrm{mod}}(h_{1,\mathrm{reg}}),
\label{eq:si_cell_growth}
\end{equation}
with
\begin{equation}
f_{\mathrm{mod}}(h_{1,\mathrm{reg}})
=
\left(1-\frac{h_u}{h_{1,\mathrm{reg}}}\right)
\left[
1-
\exp\!\left(
\frac{0.5-h_{1,\mathrm{reg}}}{0.5}
\right)
\right].
\label{eq:si_growth_modulation}
\end{equation}
Surfactant production is
\begin{equation}
R_{\Gamma}
=
k_s h_1(s_m-\Gamma)
\frac{1+\tanh[1000(s_m-\Gamma)]}{2}
\frac{1+\tanh[1000(h_{1,\mathrm{reg}}-h_u)]}{2}.
\label{eq:si_surfactant_production}
\end{equation}

The baseline aqueous-influx scale is proportional to
\begin{equation}
Q_0\frac{\phi-\phi_0}{1-\phi_0},
\label{eq:si_baseline_osmotic_source}
\end{equation}
where \(Q_0\) is the reference influx coefficient and \(\phi_0\) is
the reference cell fraction. In the inclined model, osmotic,
pressure-feedback and substrate-potential contributions are combined in
\begin{equation}
J_v
=
J_{\max}
\tanh\!\left(
\frac{J_{\mathrm{raw}}}{J_{\max}}
\right),
\label{eq:si_Jv}
\end{equation}
with
\begin{equation}
J_{\mathrm{raw}}
=
K_v
\left[
K_{\psi}(\psi-\psi_{\mathrm{eq}})
-K_p(p_g-p_{\mathrm{eq}})
+
\Lambda_{\mathrm{osm}}
\frac{\phi-\phi_0}{1-\phi_0}
\right].
\label{eq:si_Jraw}
\end{equation}
Here \(K_v\) sets the overall exchange scale, \(K_{\psi}\) and
\(K_p\) control substrate-potential and pressure feedback,
\(\Lambda_{\mathrm{osm}}\) controls osmotic exchange, and
\(J_{\max}\) bounds the flux magnitude. Positive \(J_v\)
denotes net water transfer from agar into the aqueous surface phase.

\subsubsection*{Gravity scale and geometric projection}

For a dimensional film thickness \(H\) varying over a lateral scale
\(L\), the capillary and hydrostatic pressure scales are
\begin{equation}
P_{\mathrm{cap}}
\sim
\frac{\gamma_{\mathrm{eff}}H}{L^2},
\qquad
P_g
\sim
\rho gH.
\label{eq:si_pressure_scales}
\end{equation}
Their ratio gives the estimate
\begin{equation}
G_0^{*}
\sim
\frac{P_g}{P_{\mathrm{cap}}}
=
\frac{\rho gL^2}{\gamma_{\mathrm{eff}}}.
\label{eq:si_G0_definition}
\end{equation}
Using
\(\rho=1000~\mathrm{kg\,m^{-3}}\),
\(g=9.81~\mathrm{m\,s^{-2}}\),
\(L=5~\mu\mathrm{m}\), and
\(\gamma_{\mathrm{eff}}=50~\mathrm{mN\,m^{-1}}\) gives
\begin{equation}
G_0^{*}
\approx
4.905\times10^{-6}
\approx
5.0\times10^{-6}.
\label{eq:si_G0_estimate}
\end{equation}
The characteristic length and interfacial tension were used only for
this order-of-magnitude estimate and were not independently measured in
the present SM3 experiments.

The downhill unit vector is
\begin{equation}
\mathbf e_d=-\mathbf e_y.
\label{eq:si_downhill_direction}
\end{equation}
At inclination \(\theta\), the fixed coupling is projected as
\begin{subequations}
\label{eq:si_gravity_components}
\begin{align}
G_n^{*}
&=
G_0^{*}\cos\theta,
\\
G_t^{*}
&=
G_0^{*}\sin\theta\,\mathbf e_d.
\end{align}
\end{subequations}
Here \(G_t^{*}\) denotes the surface-tangential vector coupling directed
along \(\mathbf e_d\). Thus,
\begin{equation}
(G_n^{*})^2
+
\lVert G_t^{*}\rVert^2
=
(G_0^{*})^2.
\label{eq:si_gravity_constraint}
\end{equation}

The gravity-modified film pressure and lateral driving field are
\begin{equation}
p_g
=
p_0+G_n^{*}h,
\qquad
\mathbf S
=
\nabla p_g
-
G_t^{*}.
\label{eq:si_gravity_pressure_driving}
\end{equation}
The far-field equilibrium pressure used in the vertical exchange term
is
\begin{equation}
p_{\mathrm{eq}}
=
W\left[
(h_a)^{-3}-(h_a)^{-6}
\right]
+
G_n^{*}h_a.
\label{eq:si_equilibrium_pressure}
\end{equation}
The corresponding equilibrium substrate potential is
\begin{equation}
\psi_{\mathrm{eq}}
=
p_{\mathrm{eq}}+G_n^{*}H_p.
\label{eq:si_equilibrium_potential}
\end{equation}
The auxiliary substrate-potential balance retained in the solver is
\begin{equation}
S_p\frac{\partial\psi}{\partial t}
=
\nabla\!\cdot
\left[
K_{\psi}M_{\psi}
\left(\nabla \psi-G_t^{*}\right)
\right]
+
K_{\psi}
\left[
-J_v+K_r(\psi_{\mathrm{eq}}-\psi)
\right].
\label{eq:si_substrate_potential}
\end{equation}
For every simulation reported here, \(K_{\psi}=0\). Thus,
\(\psi\) remains at its initial equilibrium value and
Eq.~\eqref{eq:si_Jraw} reduces to
\begin{equation}
J_{\mathrm{raw}}
=
K_v
\left[
-K_p(p_g-p_{\mathrm{eq}})
+
\Lambda_{\mathrm{osm}}
\frac{\phi-\phi_0}{1-\phi_0}
\right].
\label{eq:si_Jraw_reduced}
\end{equation}
The surface-tangential projection enters lateral transport directly.
The normal projection enters both the lateral pressure gradients through
\(p_g\) and the pressure-dependent vertical exchange. The
\(G_n^{*}\)-only case therefore retains both normal-pressure pathways
and does not isolate vertical replenishment alone.

\subsubsection*{Full and component-isolation simulations}

The same initial condition, parameter set, diagnostic thresholds and
final time were used in the full and component-isolation simulations.
The three cases were
\begin{subequations}
\label{eq:si_component_cases}
\begin{align}
\text{full model:}\quad
&G_n^{*}=G_0^{*}\cos\theta,
&
G_t^{*}
&=G_0^{*}\sin\theta\,\mathbf e_d,
\\
G_n^{*}\text{-only:}\quad
&G_n^{*}=G_0^{*}\cos\theta,
&
G_t^{*}
&=\mathbf 0,
\\
G_t^{*}\text{-only:}\quad
&G_n^{*}=0,
&
G_t^{*}
&=G_0^{*}\sin\theta\,\mathbf e_d.
\end{align}
\end{subequations}
Because \(\sin\theta=\sin(180^\circ-\theta)\), the \(G_t^{*}\)-only model
applies the same lateral forcing magnitude to supplementary angle
pairs. The full model distinguishes them through
\(\cos(180^\circ-\theta)=-\cos\theta\).

\subsubsection*{Initial conditions and model parameters}

The initial inoculum is represented by a smoothed disk. With
\begin{equation}
r
=
\sqrt{(x)^2+(y)^2},
\qquad
I(r)
=
\frac{1}{2}
\left[
1+
\tanh\!\left(
\frac{r_{\mathrm{in}}-r}{s_w}
\right)
\right],
\label{eq:si_initial_indicator}
\end{equation}
the initial fields are
\begin{subequations}
\label{eq:si_initial_fields}
\begin{align}
h_1
&=
h_0I
+
h_a\phi_0(1-I),
\\
h_2
&=
h_0I
+
h_a\phi_0(1-I),
\\
\Gamma
&=
s_m I+0.05s_m(1-I),
\\
\psi
&=
\psi_{\mathrm{eq}}.
\end{align}
\end{subequations}
The value \(h_0\) is assigned separately to each phase inside the
inoculum, so the initial total height there is \(2h_0\). For
\(\phi_0=0.5\), the far-field total precursor height is \(h_a\).

\begin{table*}[t]
\centering
\small
\renewcommand{\arraystretch}{1.10}
\begin{tabularx}{\textwidth}{p{0.17\textwidth} X p{0.18\textwidth}}
\toprule
Parameter & Meaning & Value \\
\midrule
\(\phi_0\) & Reference cell fraction & \(0.5\) \\
\(\beta\) & Additional cell-associated mobility coefficient & \(1.0\times10^{-10}\) \\
\(K_v\) & Overall vertical exchange coefficient; equal to the baseline influx scale \(Q_0\) & \(3.0\times10^{-4}\) \\
\(K_p\) & Pressure-feedback coefficient in vertical exchange & \(1.0\times10^{-2}\) \\
\(\Lambda_{\mathrm{osm}}\) & Osmotic-driving coefficient in vertical exchange & \(1.0\) \\
\(J_{\max}\) & Maximum magnitude of the vertical exchange flux & \(0.01\) \\
\(K_{\psi}\) & Substrate-potential coupling; inactive in reported simulations & \(0\) \\
\(M_{\psi}\) & Substrate-potential mobility in the general solver & \(1.0\times10^{-4}\) \\
\(S_p\) & Substrate-potential storage coefficient & \(1.0\) \\
\(K_r\) & Relaxation coefficient toward \(\psi_{\mathrm{eq}}\) & \(1.0\times10^{-5}\) \\
\(H_p\) & Hydrostatic offset in \(\psi_{\mathrm{eq}}\) & \(1.0\) \\
\(W\) & Disjoining-pressure/wettability coefficient & \(0.01\) \\
\(D\) & Surfactant diffusion coefficient & \(0.01\) \\
\(k_c\) & Cell-growth coefficient & \(1.0\times10^{-5}\) \\
\(k_s\) & Surfactant-production coefficient & \(1.0\times10^{-6}\) \\
\(s_m\) & Saturation surfactant concentration & \(0.1\) \\
\(r_{\mathrm{in}}\) & Initial inoculum radius & \(500\) \\
\(s_w\) & Initial transition width & \(40\) \\
\(R_{\mathcal D}\) & Radius of the circular computational domain & \(5000\) \\
\(h_0\) & Initial height assigned to each phase in the inoculum & \(4\) \\
\(h_a\) & Far-field precursor-height scale & \(1\) \\
\(h_u\) & Cell-height scale in growth and surfactant production & \(2\) \\
\(h_s\) & Height scale defining \(p_{20}=h_s\phi_0\) & \(20\) \\
\(p_{20}\) & Cell-phase carrying-capacity scale & \(10\) \\
\(G_0^{*}\) & Effective gravity-coupling strength & \(5.0\times10^{-6}\) \\
\bottomrule
\end{tabularx}
\caption{Nondimensional parameters used in the inclination simulations. The superscript \(^{*}\) is retained only on the effective gravity couplings to distinguish them from dimensional gravitational acceleration.}
\label{tab:si_model_parameters}
\end{table*}

The numerical floors and diagnostic thresholds were
\begin{equation}
h_{\min}=0.2,
\qquad
h_{1,\min}=0.05,
\qquad
h_{2,\min}=0.05,
\label{eq:si_numerical_floors}
\end{equation}
\begin{equation}
h_{1,\mathrm{thr}}=0.55,
\qquad
h_{\mathrm{thr}}=1.05,
\qquad
\epsilon_{\mathrm{diag}}=0.03.
\label{eq:si_diagnostic_thresholds}
\end{equation}

\begin{table*}[t]
\centering
\small
\renewcommand{\arraystretch}{1.08}
\begin{tabular}{ccc}
\toprule
\(\theta\) & \(G_n^{*}=G_0^{*}\cos\theta\) & \(\lVert G_t^{*}\rVert=G_0^{*}\sin\theta\) \\
\midrule
\(0^\circ\)   & \(5.0000\times10^{-6}\)  & \(0\) \\
\(30^\circ\)  & \(4.3301\times10^{-6}\)  & \(2.5000\times10^{-6}\) \\
\(45^\circ\)  & \(3.5355\times10^{-6}\)  & \(3.5355\times10^{-6}\) \\
\(75^\circ\)  & \(1.2941\times10^{-6}\)  & \(4.8296\times10^{-6}\) \\
\(105^\circ\) & \(-1.2941\times10^{-6}\) & \(4.8296\times10^{-6}\) \\
\(135^\circ\) & \(-3.5355\times10^{-6}\) & \(3.5355\times10^{-6}\) \\
\(150^\circ\) & \(-4.3301\times10^{-6}\) & \(2.5000\times10^{-6}\) \\
\(180^\circ\) & \(-5.0000\times10^{-6}\) & \(0\) \\
\bottomrule
\end{tabular}
\caption{Normal and surface-tangential projections of the fixed gravity-coupling strength. The tangential vector always points along \(\mathbf e_d=-\mathbf e_y\).}
\label{tab:si_gravity_components}
\end{table*}

\subsubsection*{Numerical implementation}

The equations were solved by finite-volume discretization on a uniform
Cartesian grid clipped by a circular domain of radius
\(R_{\mathcal D}=5000\). The inclination-sweep simulations used
\begin{equation}
N=128,
\qquad
\Delta t=10,
\qquad
t_f=4.0\times10^6,
\qquad
\Delta t_{\mathrm{out}}=1.0\times10^5,
\label{eq:si_numerical_settings}
\end{equation}
with
\begin{equation}
\Delta x=\Delta y
=
\frac{2R_{\mathcal D}}{N-1}
\approx78.74.
\label{eq:si_grid_spacing}
\end{equation}

Pressure and body-force terms were evaluated as conservative face
fluxes. For a generic scalar driving field \(\chi\), mobility
\(M\), and body-force component \(G_x\), the \(x\)-face flux
was
\begin{equation}
F_{x,i+1/2,j}^{\chi}
=
M_{i+1/2,j}
\left[
\frac{\chi_{i+1,j}-\chi_{i,j}}{\Delta x}
-
G_x
\right],
\label{eq:si_face_flux}
\end{equation}
with the analogous expression in \(y\). Mobilities on faces were
obtained by arithmetic averaging. Connections crossing the circular
mask were removed, giving a no-flux outer boundary.

Time integration used an explicit forward-Euler update with an adaptive
trial-step limiter. The accepted step was reduced when necessary so
that the maximum relative change did not exceed 0.05. The minimum
permitted step was \(10^{-6}\), at most 20 rejected trials were allowed
per update, and accepted states were required to satisfy
\(h<10^6\). After each accepted step,
\(h_1\), \(h_2\) and \(\Gamma\) were floored at
\(h_{1,\min}\), \(h_{2,\min}\) and zero, respectively.

\subsubsection*{Quantification of simulated expansion and shape}

The total wetting-front and biomass areas were calculated using smoothed
height thresholds,
\begin{subequations}
\label{eq:si_simulated_areas}
\begin{align}
A_{\mathrm{total}}(t)
&=
\int_{\mathcal D}
\frac{1+\tanh[(h-h_{\mathrm{thr}})/
\epsilon_{\mathrm{diag}}]}{2}
\,dA,
\\
A_{\mathrm{bio}}(t)
&=
\int_{\mathcal D}
\frac{1+\tanh[(h_1-h_{1,\mathrm{thr}})/
\epsilon_{\mathrm{diag}}]}{2}
\,dA.
\end{align}
\end{subequations}
The equivalent wetting radius was
\begin{equation}
R_{\mathrm{eq}}(t)
=
\sqrt{\frac{A_{\mathrm{total}}(t)}{\pi}}.
\label{eq:si_equivalent_radius}
\end{equation}
The late-stage simulated area rate was evaluated over the final quarter
of the common analysis interval,
\begin{equation}
\dot A_{\mathrm{sim}}
=
\frac{
A_{\mathrm{total}}(t_2)
-
A_{\mathrm{total}}(t_1)
}{t_2-t_1},
\qquad
t_1=0.75t_f,
\quad
t_2=t_f.
\label{eq:si_sim_late_rate}
\end{equation}
The stored outputs also contain the instantaneous finite-difference
rate
\begin{equation}
\left.\frac{dA_{\mathrm{total}}}{dt}\right|_{t_k}
=
\frac{
A_{\mathrm{total}}(t_k)
-
A_{\mathrm{total}}(t_{k-1})
}{t_k-t_{k-1}},
\label{eq:si_instantaneous_area_rate}
\end{equation}
which was used only to inspect temporal evolution.

The total-height-weighted centroid was
\begin{equation}
x_c
=
\frac{\int_{\mathcal D}xh\,dA}
{\int_{\mathcal D}h\,dA},
\qquad
y_c
=
\frac{\int_{\mathcal D}yh\,dA}
{\int_{\mathcal D}h\,dA},
\label{eq:si_sim_centroid}
\end{equation}
and its downhill displacement was
\begin{equation}
c_d
=
x_ce_{d,x}+y_ce_{d,y}.
\label{eq:si_downhill_centroid}
\end{equation}

For shape analysis, the biomass-positive region was
\begin{equation}
\mathcal B(t)
=
\left\{(x,y)\in\mathcal D:
 h_1(x,y,t)>h_{1,\mathrm{thr}}\right\}.
\label{eq:si_biomass_region}
\end{equation}
Its downhill and uphill extents were
\begin{subequations}
\label{eq:si_directional_extents}
\begin{align}
R_{\downarrow}
&=
\max_{(x,y)\in\mathcal B}
\left(xe_{d,x}+ye_{d,y}\right),
\\
R_{\uparrow}
&=
\max\left[
0,
-\min_{(x,y)\in\mathcal B}
\left(xe_{d,x}+ye_{d,y}\right)
\right].
\end{align}
\end{subequations}
These extents define
\begin{equation}
\mathcal A_{\mathrm{dir}}
=
\frac{R_{\downarrow}-R_{\uparrow}}
{R_{\downarrow}+R_{\uparrow}},
\qquad
Q_R
=
\frac{R_{\downarrow}}{R_{\uparrow}}.
\label{eq:si_directional_metrics}
\end{equation}
The same \(\mathbf e_d\) was used for full and component-isolation
runs. An additional radial diagnostic was
\begin{equation}
\mathcal A_{\mathrm{rad}}
=
\frac{R_{\max}-R_{\min}}
{R_{\max}+R_{\min}}.
\label{eq:si_radial_anisotropy}
\end{equation}
For the manuscript-level comparison, radial extents
\(R(\alpha)\) were measured from the biomass boundary and the
range-over-mean form was
\begin{equation}
A_{\mathrm{range,sim}}
=
\frac{R_{\max}-R_{\min}}{\overline R}.
\label{eq:si_sim_range_anisotropy}
\end{equation}

\subsubsection*{Sector-resolved local state near the simulated swarm edge}

Local model states were sampled from the full-model output at
\(t_f=4.0\times10^6\). The largest connected biomass-positive
component was retained, and its geometric centroid defined the sampling
origin. Five directions matched the experimental PIV sectors,
\begin{equation}
\alpha
=
\frac{\pi}{2},
\frac{3\pi}{4},
\pi,
-\frac{3\pi}{4},
-\frac{\pi}{2}.
\label{eq:si_sector_directions}
\end{equation}
For each direction, a circular ROI of radius
\(r_{\mathrm{ROI}}=3\Delta x\) was positioned one grid spacing
\(d_{\mathrm{edge}}=\Delta x\) inward from the biomass
boundary. Only ROIs lying inside the biomass-positive region were
retained. Mean \(h_1\), \(h_2\) and \(\phi\) were calculated
within each ROI.

\begin{figure*}[t]
\centering
\includegraphics[width=0.92\textwidth]
{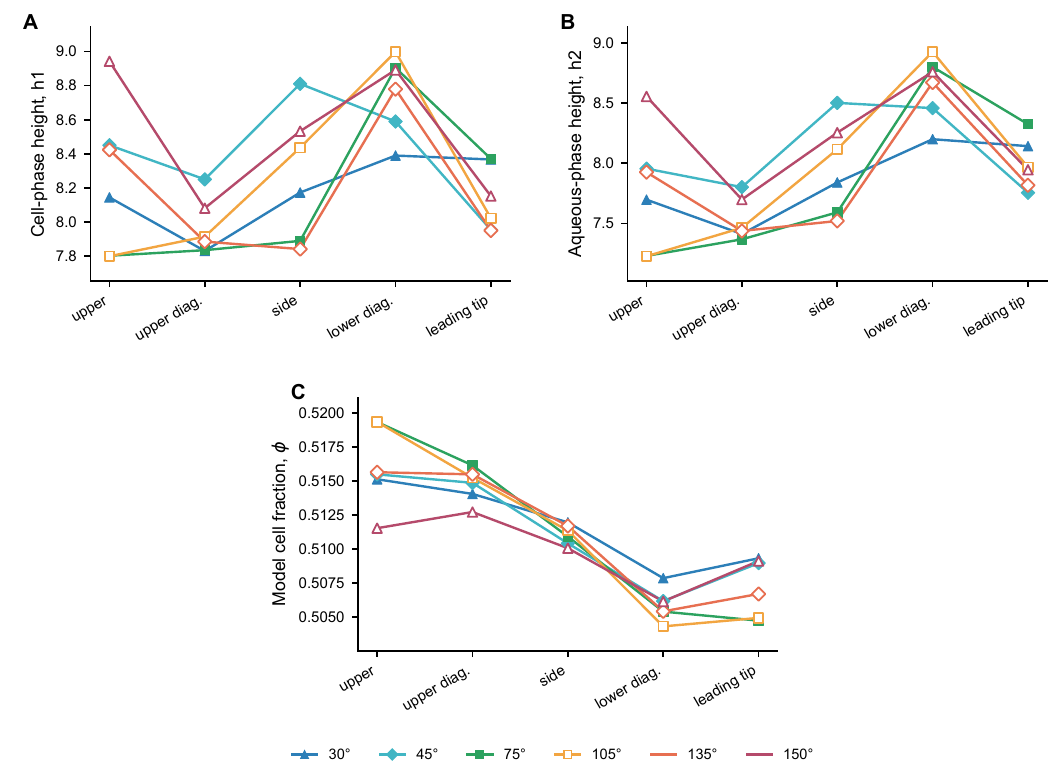}
\caption{Sector-resolved local phase-state predictions of the full
continuum model. \textbf{A}, Final-time nondimensional cell-phase height
\(h_1\). \textbf{B}, Nondimensional aqueous-phase height \(h_2\).
\textbf{C}, Model cell-fraction proxy \(\phi=h_1/(h_1+h_2)\).
Values were sampled from the five boundary-adjacent model ROIs at the
same final simulation time for the six tilted inclinations.}
\label{fig:si_local_states}
\end{figure*}

The full model predicts inclination-dependent redistribution of both
phases near the boundary. At \(75^\circ\) and \(105^\circ\), lower-side
regions were thicker than the corresponding upper region, while the
pattern weakened or reversed at other inclinations. Because \(\phi\)
often decreased as both phase heights increased, a thicker lower-side
film did not necessarily imply a larger modeled cell fraction. \FloatBarrier

\subsubsection*{Sector-resolved radial phase transport}

The solver equations have the form
\begin{equation}
\frac{\partial h_i}{\partial t}
=
\nabla\!\cdot\mathbf J_{i,\mathrm{solver}}
+
S_i.
\label{eq:si_solver_flux_form}
\end{equation}
The corresponding physical conservative flux is
\begin{equation}
\mathbf Q_i
=
-\mathbf J_{i,\mathrm{solver}}.
\label{eq:si_physical_flux_sign}
\end{equation}
Using the same sector ROIs, the radial flux was
\begin{equation}
Q_{i,r}
=
\left\langle
\mathbf Q_i\cdot\mathbf e_{\mathrm{out}}
\right\rangle_{\mathrm{ROI}},
\qquad i=1,2,
\label{eq:si_radial_phase_flux}
\end{equation}
where positive values denote outward cell-phase or aqueous-phase
transport toward the local boundary. The simulated fluxes were compared
with experimental \(v_R^{6-9}\) through relative sector ordering,
because the model was not calibrated to experimental flux units.

\begin{figure*}[t]
\centering
\includegraphics[width=0.92\textwidth]
{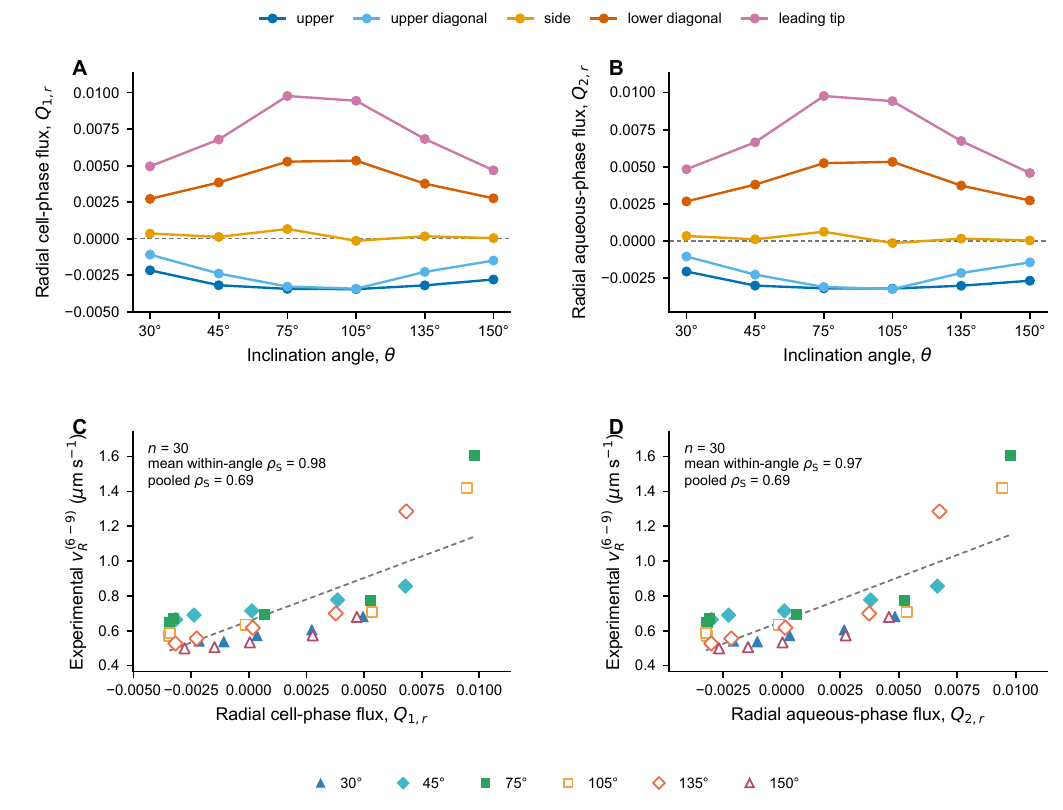}
\caption{
Sector-resolved simulated radial phase transport and its correspondence
with macroscopic directional expansion.
\textbf{A}, Nondimensional radial cell-phase flux \(Q_{1,r}\).
\textbf{B}, Nondimensional radial aqueous-phase flux \(Q_{2,r}\).
\textbf{C}, Condition-level comparison between \(Q_{1,r}\) and the
experimental late-stage directional expansion rate \(v_R^{6-9}\).
\textbf{D}, Corresponding comparison for \(Q_{2,r}\). Experimental
rates are displayed in \(\mu\mathrm{m}\,\mathrm{s}^{-1}\) after
conversion from the \(\mathrm{mm}\,\mathrm{h}^{-1}\) units used in
the main morphology figure. The 30 plotted points are angle--sector
condition means rather than independent biological replicates. The
comparison evaluates relative spatial ordering; the modeled fluxes are
coarse-grained transport quantities and are not direct measurements of
microscopic PIV velocity.
}
\label{fig:si_radial_transport}
\end{figure*}

\FloatBarrier

\subsection*{Pilot cross-species inclination assay in \textit{Bacillus subtilis}}

To examine whether inclination sensitivity was specific to
\textit{Enterobacter} sp. SM3, we reanalyzed a previously collected
pilot dataset for \textit{Bacillus subtilis}. These experiments used a
different incubation and imaging protocol and quantified colony diameter
rather than colony area. The data were therefore analyzed separately and
were not pooled with the SM3 measurements.

\begin{figure*}[t]
\centering
\includegraphics[width=0.82\textwidth]
{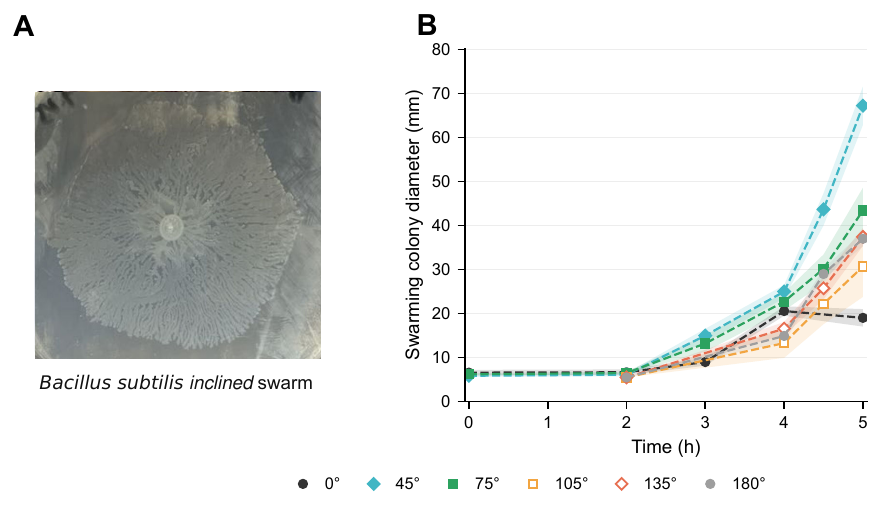}

\caption{
Exploratory inclination assay in \textit{Bacillus subtilis}.
\textbf{A}, Representative branching swarming pattern under
inclination.
\textbf{B}, Colony diameter over time for the indicated inclination
conditions. Plates were tilted after 2~h of horizontal pre-incubation.
This pilot used a different protocol and diameter-based readout from
the SM3 assay and is interpreted qualitatively.
}

\label{fig:bsubtilis_pilot_inclination}
\end{figure*}

Following 2 h of horizontal pre-incubation, colony diameter
developed an angle-dependent response. Expansion was strongest
near the intermediate acute orientation of \(45^\circ\), whereas several
larger inclinations showed weaker post-tilt diameter growth
(Supplementary Fig.~\ref{fig:bsubtilis_pilot_inclination}). The exact angular ranking differed from the SM3 dataset, but the
pilot likewise showed a non-monotonic inclination response with stronger
spreading at an intermediate acute orientation.

Unlike SM3, \textit{B. subtilis} formed branching swarming patterns,
suggesting that its macroscopic expansion is affected by additional
pattern-forming processes such as branch initiation, local wetting
heterogeneity, and colony-level morphological instability. The protocol
and diameter-based readout also differed from the main SM3 experiments,
and the replicate consistency was lower. We therefore do not treat this
dataset as a quantitative replication. The time-course data therefore suggest that inclination sensitivity
may extend beyond SM3, while the differences in protocol, morphology and
replicate consistency preclude a quantitative cross-species comparison.

\FloatBarrier

\FloatBarrier




\begin{thebibliography}{40}
\providecommand{\natexlab}[1]{#1}
\providecommand{\url}[1]{\texttt{#1}}
\expandafter\ifx\csname urlstyle\endcsname\relax
  \providecommand{\doi}[1]{doi: #1}\else
  \providecommand{\doi}{doi: \begingroup \urlstyle{rm}\Url}\fi

\bibitem[Harshey(2003)]{harshey_surface_2003}
Rasika~M. Harshey.
\newblock Bacterial motility on a surface: Many ways to a common goal.
\newblock \emph{Annual Review of Microbiology}, 57:\penalty0 249--273, 2003.
\newblock \doi{10.1146/annurev.micro.57.030502.091014}.

\bibitem[Kearns(2010)]{kearns_field_2010}
Daniel~B. Kearns.
\newblock A field guide to bacterial swarming motility.
\newblock \emph{Nature Reviews Microbiology}, 8\penalty0 (9):\penalty0
  634--644, 2010.
\newblock \doi{10.1038/nrmicro2405}.

\bibitem[Partridge(2022)]{partridge_surveying_2022}
Jonathan~D. Partridge.
\newblock Surveying a swarm: Experimental techniques to establish and examine
  bacterial collective motion.
\newblock \emph{Applied and Environmental Microbiology}, 88\penalty0
  (3):\penalty0 e01853--21, 2022.
\newblock \doi{10.1128/AEM.01853-21}.

\bibitem[Darnton et~al.(2010)Darnton, Turner, Rojevsky, and
  Berg]{darnton_dynamics_2010}
Nicholas~C. Darnton, Linda Turner, Svetlana Rojevsky, and Howard~C. Berg.
\newblock Dynamics of bacterial swarming.
\newblock \emph{Biophysical Journal}, 98\penalty0 (10):\penalty0 2082--2090,
  2010.
\newblock \doi{10.1016/j.bpj.2010.01.053}.

\bibitem[Zhang et~al.(2010)Zhang, Be'er, Florin, and
  Swinney]{zhang_collective_2010}
H.~P. Zhang, Avraham Be'er, E.-L. Florin, and Harry~L. Swinney.
\newblock Collective motion and density fluctuations in bacterial colonies.
\newblock \emph{Proceedings of the National Academy of Sciences of the United
  States of America}, 107\penalty0 (31):\penalty0 13626--13630, 2010.
\newblock \doi{10.1073/pnas.1001651107}.

\bibitem[Be'er and Ariel(2019)]{beer_statphys_2019}
Avraham Be'er and Gil Ariel.
\newblock A statistical physics view of swarming bacteria.
\newblock \emph{Movement Ecology}, 7\penalty0 (1):\penalty0 9, 2019.
\newblock \doi{10.1186/s40462-019-0153-9}.

\bibitem[Aranson(2022)]{aranson_bacterial_2022}
Igor~S. Aranson.
\newblock Bacterial active matter.
\newblock \emph{Reports on Progress in Physics}, 85\penalty0 (7):\penalty0
  076601, 2022.
\newblock \doi{10.1088/1361-6633/ac723d}.

\bibitem[Copeland and Weibel(2009)]{copeland_bacterial_2009}
Matthew~F. Copeland and Douglas~B. Weibel.
\newblock Bacterial swarming: a model system for studying dynamic
  self-assembly.
\newblock \emph{Soft Matter}, 5\penalty0 (6):\penalty0 1174--1187, 2009.
\newblock \doi{10.1039/B812146J}.

\bibitem[Trinschek et~al.(2018)Trinschek, John, and
  Thiele]{trinschek_front_2018}
Sarah Trinschek, Karin John, and Uwe Thiele.
\newblock Modelling of surfactant-driven front instabilities in spreading
  bacterial colonies.
\newblock \emph{Soft Matter}, 14\penalty0 (22):\penalty0 4464--4476, 2018.
\newblock \doi{10.1039/C8SM00422F}.

\bibitem[Srinivasan et~al.(2019)Srinivasan, Kaplan, and
  Mahadevan]{srinivasan_multiphase_2019}
Siddarth Srinivasan, C.~Nadir Kaplan, and L.~Mahadevan.
\newblock A multiphase theory for spreading microbial swarms and films.
\newblock \emph{eLife}, 8:\penalty0 e42697, 2019.
\newblock \doi{10.7554/eLife.42697}.

\bibitem[Kotian et~al.(2020)Kotian, Abdulla, Hithysini, Harkar, Joge, Mishra,
  Singh, and Varma]{PhysRevE.101.012407}
Harshitha~S. Kotian, Amith~Z. Abdulla, K.~N. Hithysini, Shalini Harkar, Shubham
  Joge, Ayushi Mishra, Varsha Singh, and Manoj~M. Varma.
\newblock Active modulation of surfactant-driven flow instabilities by swarming
  bacteria.
\newblock \emph{Phys. Rev. E}, 101:\penalty0 012407, Jan 2020.
\newblock \doi{10.1103/PhysRevE.101.012407}.
\newblock URL \url{https://link.aps.org/doi/10.1103/PhysRevE.101.012407}.

\bibitem[Kumar and Pushpavanam(2025)]{kumar_generalized_2025}
Uttam Kumar and Subramaniam Pushpavanam.
\newblock A generalized model for predicting different morphologies of
  bacterial swarming on a porous solid surface.
\newblock \emph{Soft Matter}, 21\penalty0 (2):\penalty0 226--240, 2025.
\newblock \doi{10.1039/D4SM01072H}.

\bibitem[Marchetti et~al.(2013)Marchetti, Joanny, Ramaswamy, Liverpool, Prost,
  Rao, and Simha]{marchetti_hydrodynamics_2013}
M.~C. Marchetti, Jean-Fran{\c{c}}ois Joanny, Sriram Ramaswamy, T.~B. Liverpool,
  Jacques Prost, Madan Rao, and R.~Aditi Simha.
\newblock Hydrodynamics of soft active matter.
\newblock \emph{Reviews of Modern Physics}, 85\penalty0 (3):\penalty0
  1143--1189, 2013.
\newblock \doi{10.1103/RevModPhys.85.1143}.

\bibitem[Wu et~al.(2011)Wu, Hosu, and Berg]{wu_microbubbles_2011}
Yilin Wu, Basarab~G. Hosu, and Howard~C. Berg.
\newblock Microbubbles reveal chiral fluid flows in bacterial swarms.
\newblock \emph{Proceedings of the National Academy of Sciences of the United
  States of America}, 108\penalty0 (10):\penalty0 4147--4151, 2011.
\newblock \doi{10.1073/pnas.1016693108}.

\bibitem[Wu and Berg(2012)]{wu_water_2012}
Yilin Wu and Howard~C. Berg.
\newblock Water reservoir maintained by cell growth fuels the spreading of a
  bacterial swarm.
\newblock \emph{Proceedings of the National Academy of Sciences of the United
  States of America}, 109\penalty0 (11):\penalty0 4128--4133, 2012.
\newblock \doi{10.1073/pnas.1118238109}.

\bibitem[Ping et~al.(2014)Ping, Wu, Hosu, Tang, and Berg]{ping_osmotic_2014}
Liyan Ping, Yilin Wu, Basarab~G. Hosu, Jay~X. Tang, and Howard~C. Berg.
\newblock Osmotic pressure in a bacterial swarm.
\newblock \emph{Biophysical Journal}, 107\penalty0 (4):\penalty0 871--878,
  2014.
\newblock \doi{10.1016/j.bpj.2014.05.052}.

\bibitem[Angelini et~al.(2009)Angelini, Roper, Kolter, Weitz, and
  Brenner]{angelini_bacillus_2009}
Thomas~E. Angelini, Marcus Roper, Roberto Kolter, David~A. Weitz, and
  Michael~P. Brenner.
\newblock {Bacillus subtilis} spreads by surfing on waves of surfactant.
\newblock \emph{Proceedings of the National Academy of Sciences of the United
  States of America}, 106\penalty0 (43):\penalty0 18109--18113, 2009.
\newblock \doi{10.1073/pnas.0905890106}.

\bibitem[Fauvart et~al.(2012)Fauvart, Phillips, Bachaspatimayum, Verstraeten,
  Fransaer, Michiels, and Vermant]{fauvart_surface_2012}
Maarten Fauvart, P.~Phillips, D.~Bachaspatimayum, Natalie Verstraeten, Jan
  Fransaer, Jan Michiels, and Jan Vermant.
\newblock Surface tension gradient control of bacterial swarming in colonies of
  {Pseudomonas aeruginosa}.
\newblock \emph{Soft Matter}, 8\penalty0 (1):\penalty0 70--76, 2012.
\newblock \doi{10.1039/C1SM06002C}.

\bibitem[Ke et~al.(2015)Ke, Hsueh, Cheng, Wu, and Liu]{ke_water_2015}
Wan-Ju Ke, Yi-Huang Hsueh, Yu-Chieh Cheng, Chih-Ching Wu, and Shih-Tung Liu.
\newblock Water surface tension modulates the swarming mechanics of {Bacillus
  subtilis}.
\newblock \emph{Frontiers in Microbiology}, 6:\penalty0 1017, 2015.
\newblock \doi{10.3389/fmicb.2015.01017}.

\bibitem[Wilkins et~al.(1969)Wilkins, Hatten, and Salmirs]{wilkins_effect_1969}
Judd~R. Wilkins, Mary~B. Hatten, and Seymour Salmirs.
\newblock Effect of gravity on the colonial morphology of staphylococci in soft
  agar.
\newblock \emph{Applied Microbiology}, 18\penalty0 (4):\penalty0 680--681,
  1969.
\newblock \doi{10.1128/am.18.4.680-681.1969}.

\bibitem[Marra et~al.(2025)Marra, Rizzo, and Caserta]{marra_microfluidics_2025}
Daniele Marra, Moreno Rizzo, and Sergio Caserta.
\newblock Microfluidics unveils role of gravity and shear stress on
  {Pseudomonas fluorescens} motility and biofilm growth.
\newblock \emph{npj Biofilms and Microbiomes}, 11:\penalty0 122, 2025.
\newblock \doi{10.1038/s41522-025-00744-4}.

\bibitem[Malea et~al.(2025)Malea, Petala, Kostoglou, and
  Karapantsios]{malea_contribution_2025}
Elisavet Malea, Maria Petala, Margaritis Kostoglou, and Theodoros Karapantsios.
\newblock Contribution of the gravity component and surface type during the
  initial stages of biofilm formation at solid--liquid interfaces.
\newblock \emph{Water}, 17\penalty0 (15):\penalty0 2277, 2025.
\newblock \doi{10.3390/w17152277}.

\bibitem[Volpe et~al.(2022)Volpe, Bechinger, Cichos, Golestanian, L{\"o}wen,
  Sperl, and Volpe]{volpe_active_2022}
Giorgio Volpe, Clemens Bechinger, Frank Cichos, Ramin Golestanian, Hartmut
  L{\"o}wen, Matthias Sperl, and Giovanni Volpe.
\newblock Active matter in space.
\newblock \emph{npj Microgravity}, 8:\penalty0 54, 2022.
\newblock \doi{10.1038/s41526-022-00230-7}.

\bibitem[Hennes et~al.(2017)Hennes, Tailleur, Charron, and
  Daerr]{hennes_active_2017}
Marc Hennes, Julien Tailleur, Ga{\"e}lle Charron, and Adrian Daerr.
\newblock Active depinning of bacterial droplets: the collective surfing of
  {Bacillus subtilis}.
\newblock \emph{Proceedings of the National Academy of Sciences of the United
  States of America}, 114\penalty0 (23):\penalty0 5958--5963, 2017.
\newblock \doi{10.1073/pnas.1703997114}.

\bibitem[Chen et~al.(2021)Chen, Mani, Karani, Li, Mani, and
  Tang]{chen_confinement_2021}
Weijie Chen, Neha Mani, Hamid Karani, Hao Li, Sridhar Mani, and Jay~X. Tang.
\newblock Confinement discerns swarmers from planktonic bacteria.
\newblock \emph{eLife}, 10:\penalty0 e64176, 2021.
\newblock \doi{10.7554/eLife.64176}.

\bibitem[Thielicke and Sonntag(2021)]{thielicke_pivlab_2021}
William Thielicke and Ren{\'e} Sonntag.
\newblock Particle image velocimetry for {MATLAB}: Accuracy and enhanced
  algorithms in {PIVlab}.
\newblock \emph{Journal of Open Research Software}, 9\penalty0 (1):\penalty0
  12, 2021.
\newblock \doi{10.5334/jors.334}.

\bibitem[Graftieaux et~al.(2001)Graftieaux, Michard, and
  Grosjean]{graftieaux_combining_2001}
Laurent Graftieaux, Marc Michard, and Nathalie Grosjean.
\newblock Combining {PIV}, {POD} and vortex identification algorithms for the
  study of unsteady turbulent swirling flows.
\newblock \emph{Measurement Science and Technology}, 12\penalty0 (9):\penalty0
  1422--1429, 2001.
\newblock \doi{10.1088/0957-0233/12/9/307}.

\bibitem[Seminara et~al.(2012)Seminara, Angelini, Wilking, Vlamakis, Ebrahim,
  Kolter, Weitz, and Brenner]{seminara_osmotic_2012}
Agnese Seminara, Thomas~E. Angelini, James~N. Wilking, Hera Vlamakis, Senan
  Ebrahim, Roberto Kolter, David~A. Weitz, and Michael~P. Brenner.
\newblock Osmotic spreading of bacillus subtilis biofilms driven by an
  extracellular matrix.
\newblock \emph{Proceedings of the National Academy of Sciences of the United
  States of America}, 109\penalty0 (4):\penalty0 1116--1121, 2012.
\newblock \doi{10.1073/pnas.1109261108}.

\bibitem[Trinschek et~al.(2016)Trinschek, John, and
  Thiele]{trinschek_osmotic_2016}
Sarah Trinschek, Karin John, and Uwe Thiele.
\newblock From a thin film model for passive suspensions towards the
  description of osmotic biofilm spreading.
\newblock \emph{AIMS Materials Science}, 3\penalty0 (3):\penalty0 1138--1159,
  2016.
\newblock \doi{10.3934/matersci.2016.3.1138}.

\bibitem[Dechesne et~al.(2010)Dechesne, Wang, G{\"u}lez, Or, and
  Smets]{dechesne_hydration_2010}
Arnaud Dechesne, Gang Wang, Gamze G{\"u}lez, Dani Or, and Barth~F. Smets.
\newblock Hydration-controlled bacterial motility and dispersal on surfaces.
\newblock \emph{Proceedings of the National Academy of Sciences of the United
  States of America}, 107\penalty0 (32):\penalty0 14369--14372, 2010.
\newblock \doi{10.1073/pnas.1008392107}.

\bibitem[Tecon and Or(2016)]{tecon_film_2016}
Robin Tecon and Dani Or.
\newblock Bacterial flagellar motility on hydrated rough surfaces controlled by
  aqueous film thickness and connectedness.
\newblock \emph{Scientific Reports}, 6:\penalty0 19409, 2016.
\newblock \doi{10.1038/srep19409}.

\bibitem[Ilkanaiv et~al.(2017)Ilkanaiv, Kearns, Ariel, and
  {Be'er}]{ilkanaiv_aspect_2017}
Bella Ilkanaiv, Daniel~B. Kearns, Gil Ariel, and Avraham {Be'er}.
\newblock Effect of cell aspect ratio on swarming bacteria.
\newblock \emph{Physical Review Letters}, 118:\penalty0 158002, 2017.
\newblock \doi{10.1103/PhysRevLett.118.158002}.

\bibitem[Be'er et~al.(2020)Be'er, Ilkanaiv, Gross, Kearns, Heidenreich,
  B{\"a}r, and Ariel]{beer_phase_2020}
Avraham Be'er, Bella Ilkanaiv, Renan Gross, Daniel~B. Kearns, Sebastian
  Heidenreich, Markus B{\"a}r, and Gil Ariel.
\newblock A phase diagram for bacterial swarming.
\newblock \emph{Communications Physics}, 3\penalty0 (1):\penalty0 66, 2020.
\newblock \doi{10.1038/s42005-020-0327-1}.
\newblock URL \url{https://doi.org/10.1038/s42005-020-0327-1}.

\bibitem[Wessel et~al.(2014)Wessel, Arshad, Fitzpatrick, Connell, Bonnecaze,
  Shear, and Whiteley]{wessel_oxygen_2014}
Aimee~K. Wessel, Talha~A. Arshad, Mignon Fitzpatrick, Jodi~L. Connell, Roger~T.
  Bonnecaze, Jason~B. Shear, and Marvin Whiteley.
\newblock Oxygen limitation within a bacterial aggregate.
\newblock \emph{mBio}, 5\penalty0 (2):\penalty0 e00992--14, 2014.
\newblock \doi{10.1128/mBio.00992-14}.

\bibitem[Partridge et~al.(2018)Partridge, Ariel, Schvartz, Harshey, and
  {Be'er}]{partridge_architecture_2018}
Jonathan~D. Partridge, Gil Ariel, Orly Schvartz, Rasika~M. Harshey, and Avraham
  {Be'er}.
\newblock The 3d architecture of a bacterial swarm has implications for
  antibiotic tolerance.
\newblock \emph{Scientific Reports}, 8:\penalty0 15823, 2018.
\newblock \doi{10.1038/s41598-018-34192-2}.

\bibitem[Be'er et~al.(2009)Be'er, Smith, Zhang, Florin, Payne, and
  Swinney]{beer_surfactant_growth_2009}
Avraham Be'er, Rachel~S. Smith, H.~P. Zhang, E.-L. Florin, Shelley~M. Payne,
  and Harry~L. Swinney.
\newblock {Paenibacillus dendritiformis} bacterial colony growth depends on
  surfactant but not on bacterial motion.
\newblock \emph{Journal of Bacteriology}, 191\penalty0 (18):\penalty0
  5758--5764, 2009.
\newblock \doi{10.1128/JB.00660-09}.

\bibitem[Xu et~al.(2019)Xu, Dauparas, Das, Lauga, and
  Wu]{xu_selforganization_2019}
Haoran Xu, Justas Dauparas, Debasish Das, Eric Lauga, and Yilin Wu.
\newblock Self-organization of swimmers drives long-range fluid transport in
  bacterial colonies.
\newblock \emph{Nature Communications}, 10:\penalty0 1792, 2019.
\newblock \doi{10.1038/s41467-019-09818-2}.

\bibitem[Li et~al.(2024)Li, Chat{\'e}, Sano, Shi, and
  Zhang]{li_robust_edge_2024}
He~Li, Hugues Chat{\'e}, Masaki Sano, Xia-qing Shi, and H.~P. Zhang.
\newblock Robust edge flows in swarming bacterial colonies.
\newblock \emph{Physical Review X}, 14:\penalty0 041006, 2024.
\newblock \doi{10.1103/PhysRevX.14.041006}.

\bibitem[Xu et~al.(2023)Xu, Nejad, Yeomans, and Wu]{xu_invasion_2023}
Haoran Xu, Mehrana~R. Nejad, Julia~M. Yeomans, and Yilin Wu.
\newblock Geometrical control of interface patterning underlies active matter
  invasion.
\newblock \emph{Proceedings of the National Academy of Sciences}, 120\penalty0
  (30):\penalty0 e2219708120, 2023.
\newblock \doi{10.1073/pnas.2219708120}.

\bibitem[Patteson et~al.(2018)Patteson, Gopinath, and
  Arratia]{patteson_interface_2018}
Alison~E. Patteson, Arvind Gopinath, and Paulo~E. Arratia.
\newblock The propagation of active-passive interfaces in bacterial swarms.
\newblock \emph{Nature Communications}, 9:\penalty0 5373, 2018.
\newblock \doi{10.1038/s41467-018-07781-y}.

\end{thebibliography}

\begin{thebibliography}{2}
\providecommand{\natexlab}[1]{#1}
\providecommand{\url}[1]{\texttt{#1}}
\expandafter\ifx\csname urlstyle\endcsname\relax
  \providecommand{\doi}[1]{doi: #1}\else
  \providecommand{\doi}{doi: \begingroup \urlstyle{rm}\Url}\fi

\bibitem[Graftieaux et~al.(2001)Graftieaux, Michard, and
  Grosjean]{SIREF-graftieaux_combining_2001}
Laurent Graftieaux, Marc Michard, and Nathalie Grosjean.
\newblock Combining {PIV}, {POD} and vortex identification algorithms for the
  study of unsteady turbulent swirling flows.
\newblock \emph{Measurement Science and Technology}, 12\penalty0 (9):\penalty0
  1422--1429, 2001.
\newblock \doi{10.1088/0957-0233/12/9/307}.

\bibitem[Kumar and Pushpavanam(2025)]{SIREF-kumar_generalized_2025}
Uttam Kumar and Subramaniam Pushpavanam.
\newblock A generalized model for predicting different morphologies of
  bacterial swarming on a porous solid surface.
\newblock \emph{Soft Matter}, 21\penalty0 (2):\penalty0 226--240, 2025.
\newblock \doi{10.1039/D4SM01072H}.

\end{thebibliography}
\end{document}